\documentclass[12pt]{iopart}
\usepackage[pdftex]{graphicx}
\begin{document}

\title {FDTD Computation of Human Eye Exposure 
to Ultra-wideband Electromagnetic Pulses}

\author{Neven Simicevic \footnote[3]{Correspondence should be addressed to Louisiana Tech University, 
PO Box 10348, Ruston, LA 71272, Tel: +1.318.257.3591, Fax: +1.318.257.4228, 
E-mail: neven@phys.latech.edu}}

\address{\ Center for Applied Physics Studies, Louisiana Tech University,
 Ruston, LA 71272, USA}

\begin{abstract}

With an increase in the application of ultra-wideband (UWB) electromagnetic pulses 
in the communications industry, radar, biotechnology and medicine, comes an interest in
UWB exposure safety standards. Despite an increase of  the scientific research on 
bioeffects of exposure to non-ionizing UWB pulses, characterization of those effects is far 
from complete. A numerical computational approach, such as a  finite-difference time domain 
(FDTD) method, is required to visualize and  understand the complexity of broadband 
electromagnetic interactions. The FDTD method has almost no 
limits in the description of the geometrical and dispersive properties of the simulated material, 
it is numerically robust and appropriate for current computer technology.
In this paper, a complete calculation of exposure of the human eye to UWB 
electromagnetic pulses in the frequency range of 3.1-10.6, 22-29, and 
57-64  GHz is performed. Computation in this frequency range required a geometrical
resolution of the eye of $\rm 0.1 \: mm$ and an arbitrary precision in the description of 
its dielectric properties in terms of the Debye model. New results show that the
interaction of UWB pulses with the eye tissues exhibits the same properties as 
the interaction of the continuous electromagnetic waves (CW) with the frequencies
from the pulse's  frequency spectrum. It is also shown that under the same exposure
conditions the exposure to UWB pulses is from one to many orders of magnitude 
safer than the exposure to CW.

\end{abstract}



\maketitle

\section{Introduction}

While a large number of experiments have been performed in an attempt 
to provide an insight into the biological effects of the electromagnetic fields 
(Barnes and Greenebaum 2006, Miller {\it et al} 2002, Polk and Postow 1995),  
the bioeffects of non-ionizing ultra-wideband (UWB) electromagnetic (EM) 
pulses have not been studied in as much detail as the effects of continuous-wave 
(CW) radiation. Nevertheless, as the interest for the application of UWB electromagnetic pulses 
increases, particularly in the communications industry and medicine, so does the interest 
to understand their bioeffects (Zastrow {\it et al} 2007, 
Ji {\it et al} 2006, Hu {\it et al} 2005, Schoenbach {\it et al} 2004). 
Only recently was the work in the UWB field in the United States declassified (Taylor 1995), 
and only in 2002 were  UWB pulses approved 
by the Federal Communications Commission  in the U.S. for ``applications such as radar
imaging of objects buried under the ground or behind walls and short-range, 
high-speed data transmissions" (FCC 2002).

A computational approach to the exposure of biological tissues to non-ionizing 
UWB radiation is more involved than the computation of exposure to CW radiation.
In addition to a realistic description of the geometry, the 
physical properties of exposed biological material have to be known over a broad frequency range.
The computation has to be numerically robust and appropriate for the computer technology of today. 
The  finite difference-time domain (FDTD) method, introduced by Kane Yee in the 1960s (Yee 1966)
and extensively developed in the 1990s (Sadiku 1992, Kunz and Luebbers 1993, 
Sullivan 2000, Taflove and Hagness 2000), is a well known numerical method that satisfies these 
conditions. 
 
The response of a biological system to an EM pulse relates to the extent of conversion of 
EM pulse energy into mechanical or thermal energy as the pulse propagates through the material.
Calculating the electric and magnetic fields around and inside the 
exposed material and combining these with the electromagnetic properties of 
the exposed material, the specific absorption rate, total deposited energy, and induced current inside 
the sample can be calculated. Finally, the interaction mechanism between the 
electromagnetic radiation and biological material can be modeled and understood. 
It is possible that the bioeffects of short EM pulses are qualitatively different from those 
of narrow-band radio frequencies. If, for example, the specific absorption rate 
standards defined for continuous radio frequency by the American National Standard Institute
(ANSI 1992) or by the International Commission on Non-Ionizing Radiation Protection
(ICNIRP 1998) are applied to the short UWB pulses, their meaning loses its clarity. 
Because of the pulses' short duration and as a result of the non-uniform power absorption, 
the deposition power and the induced current densities are large and can exceed locally
even the highest allowed power limit of 4 W/kg (Simicevic 2007).
At the same time, the energy from the exposure to an UWB pulse is too small to induce 
an increase in the temperature of the exposed sample due to, compared to CW, the much lower 
energy density of the UWB pulse.

We have already applied the FDTD method  to calculate the EM fields inside biological 
samples exposed to nanopulses in a GTEM cell (Simicevic and Haynie 2005, Simicevic 2005). 
We included in the calculation the geometry and the physical properties of 
the samples to the fullest extent and applied a full three-dimensional 
computation, but have restricted the calculation to one sample (blood) and the 
frequency range up to $\rm 15 \; GHz$.

In this paper, the effects of the exposure of a human eye to UWB radiation 
in the frequency range of up to $\rm 90 \; GHz$ are calculated using the same three-dimensional  
FDTD computer code described and validated in the previous works 
(Simicevic and Haynie 2005, Simicevic 2005). The shape and the physical properties of
the eye are described as accurately as allowed by the existing data. The computational
space of $\rm 30 \; mm$ $\times$ $\rm 26 \; mm$ $\times$ $\rm 26 \; mm$ was divided 
into $\rm \sim 2.0 \times 10^{7}$ cells and the computation was performed at the Louisiana 
Optical Initiative (LONI) cluster of supercomputers. The eye was exposed to the 
UWB pulses from three different frequency regions: $\rm 3.1 - 10.6 \; GHz$ 
(authorized for communications and radar imaging), 
$\rm 22 - 29 \; GHz$ (authorized for vehicle radars), and $\rm 57 - 64 \; GHz$
(authorized for unlicensed use)  (FCC 2002).

What follows is a detailed description of the computation: geometrical and physical 
properties of the eye, the properties of the UWB pulses, the details of the computation
and data representation, and the results.

\section{The Shape and the Size of a Human Eye}

\begin{figure}
\begin{center}
\includegraphics{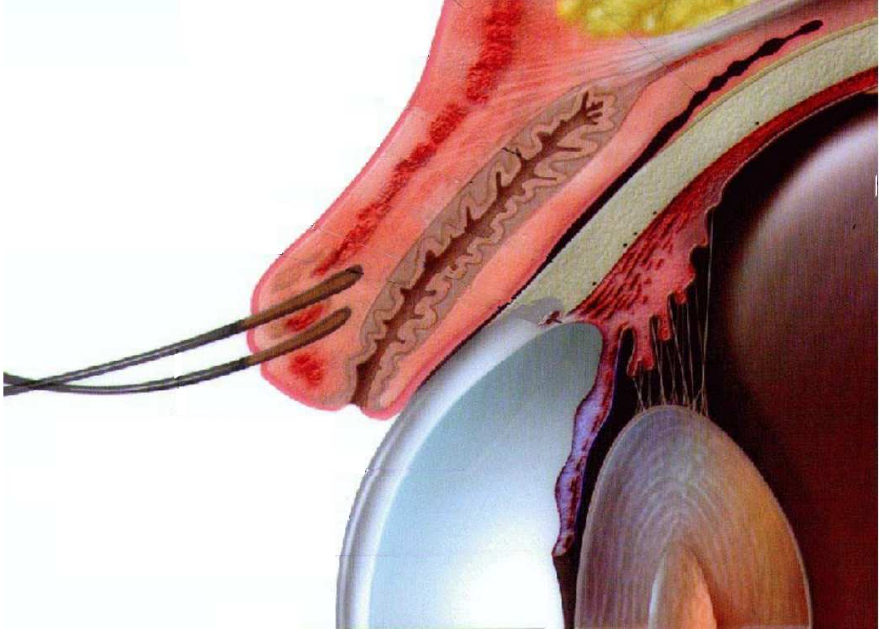}
\caption{\label{eye1} Schematics of the human eye.}
\end{center}
\end{figure}

The shape and the size of a human eye, an example of which is shown  
in Figure~\ref{eye1}, varies from one individual to another and changes with age.  
While the theoretical eye models put the emphasis mostly on the optical properties of the eye 
(Norrby {\it et al} 2007, de AlmeidaI and Carvalho 2007, Siedlecki {\it et al} 2004, Lotmar 1971), 
they can as well be used in radiation protection (Charles and Brown 1975).

\begin{figure}
\begin{center}
\includegraphics[scale=0.5]{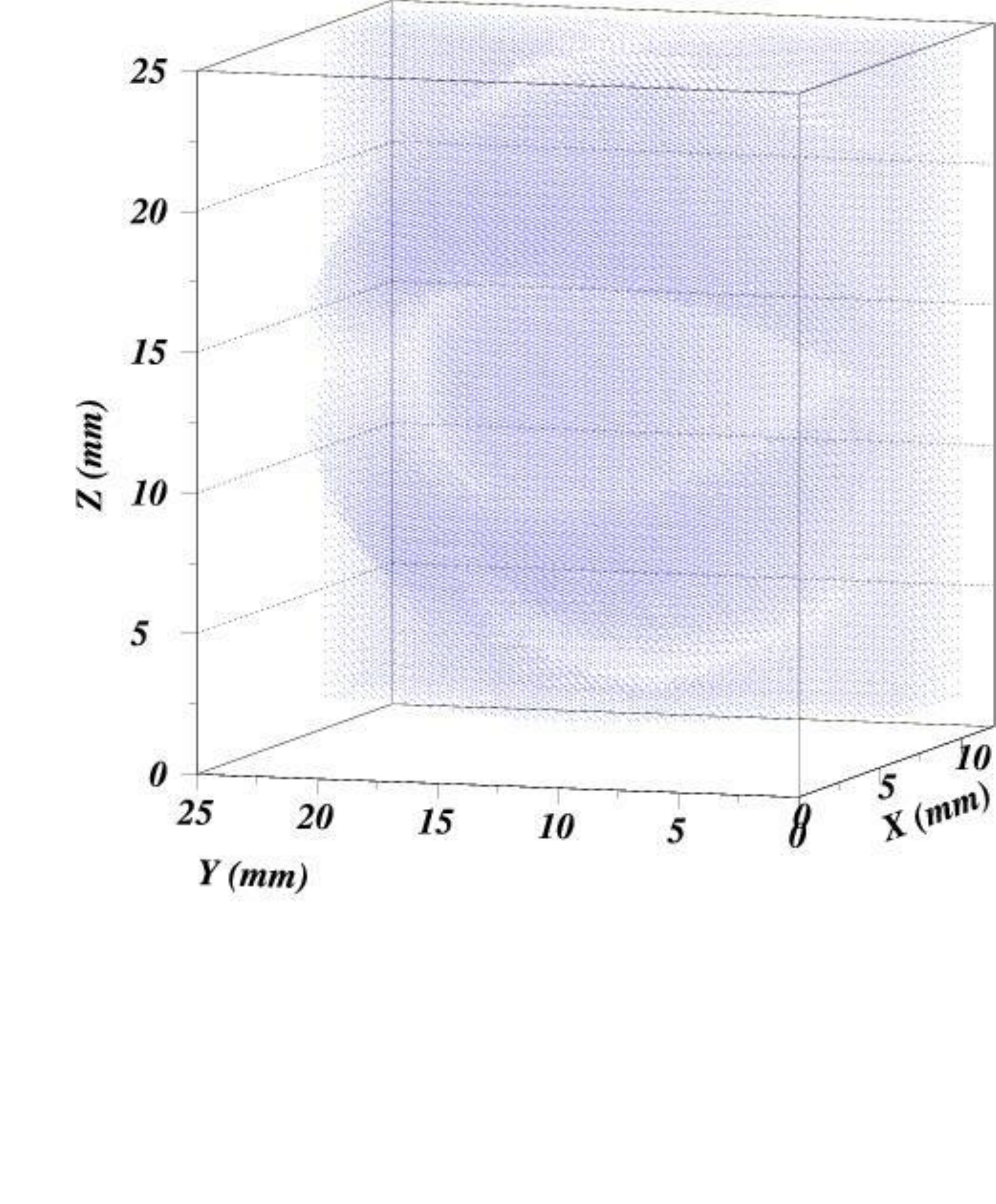}
\vspace*{- 2.5cm}
\caption{\label{eye_front} Front of the human eye discretized by means of Yee cells. 
One cell in the figure represents 16 Yee cells used in the computation.}
\end{center}
\end{figure}

For the application in the FDTD computing, the shape of the human eye was discretized by 
means of Yee cells, cubes of an edge length $\Delta x$ (Yee 1966). The two-dimensional 
shape of the eye shown in Figure~\ref{eye1} was normalized to the dimensions of the 
existing theoretical eye models, the pixels belonging to different eye 
tissues were retrieved by image processing software, and, because of rotational symmetry,
rotation around the proper axis of rotation was performed to create the three dimensional model  of the eye. 
To account for as small structures of the eye as, for example, the thickness of the cornea 
(only $\rm \sim 0.6 - 0.7 \; mm$ in size),  and still be able to perform the computation in a reasonable time, the  
lengths of the Yee cell were chosen to be $\rm \Delta x = \Delta y = \Delta z = 0.1  \; mm$
The computation was performed inside the computational space of 
$\rm 30 \; mm$ $\times$ $\rm 26 \; mm$ $\times$ $\rm 26 \; mm$ in size 
discretized by $\rm \sim 2.0 \times 10^{7}$ cells. 
The front part of the eye constructed from Yee cells is shown in Figure~\ref{eye_front}.
To make Figure~\ref{eye_front} readable, 16 Yee cells used in the computation are presented 
as one cell in the  figure.

During the process of discretization of the human eye, the structures of the eye were separated into
the tissues of known dielectric properties:  sclera, vitreous humour, lens cortex, cornea, 
lens nucleus, muscles, and blood. The eye was embedded in the bone, muscle and skin tissue 
(wet and dry).   When dealing with UWB radiation it was crucial  to have a proper description 
of the dielectric properties of the exposed tissues over the entire 
frequency range. They were taken from the established references  
(Gabriel 1996, Gabriel C {\it et al} 1996, Gabriel S {\it et al} 1996).

From the relation between the size of a Yee cell $\Delta x$ and the maximum 
frequency $f_{max}$ allowed in the FDTD computation (Kunz and Luebbers 1993, 
Taflove and Hagness 2000)

\begin{equation}
\Delta x \simeq {v \over {10 \; f_{max}}},
\end{equation}
where $v$ is the speed of light in the material, the Yee cube edge length of 
$\rm \Delta x = 0.1 \; mm$ may be used to model exposure to the wave frequency 
of up to $\rm  90 \; GHz$. This value is far greater than the frequency range of the UWB pulses
used in the calculation. We have achieved an optimal agreement between 
geometrical description of the eye and the computational requirements 
of the FDTD method.

\section{Shape of the UWB Pulses}

\begin{figure}
\begin{center}
\hspace*{-1.cm}\includegraphics[scale=0.33]{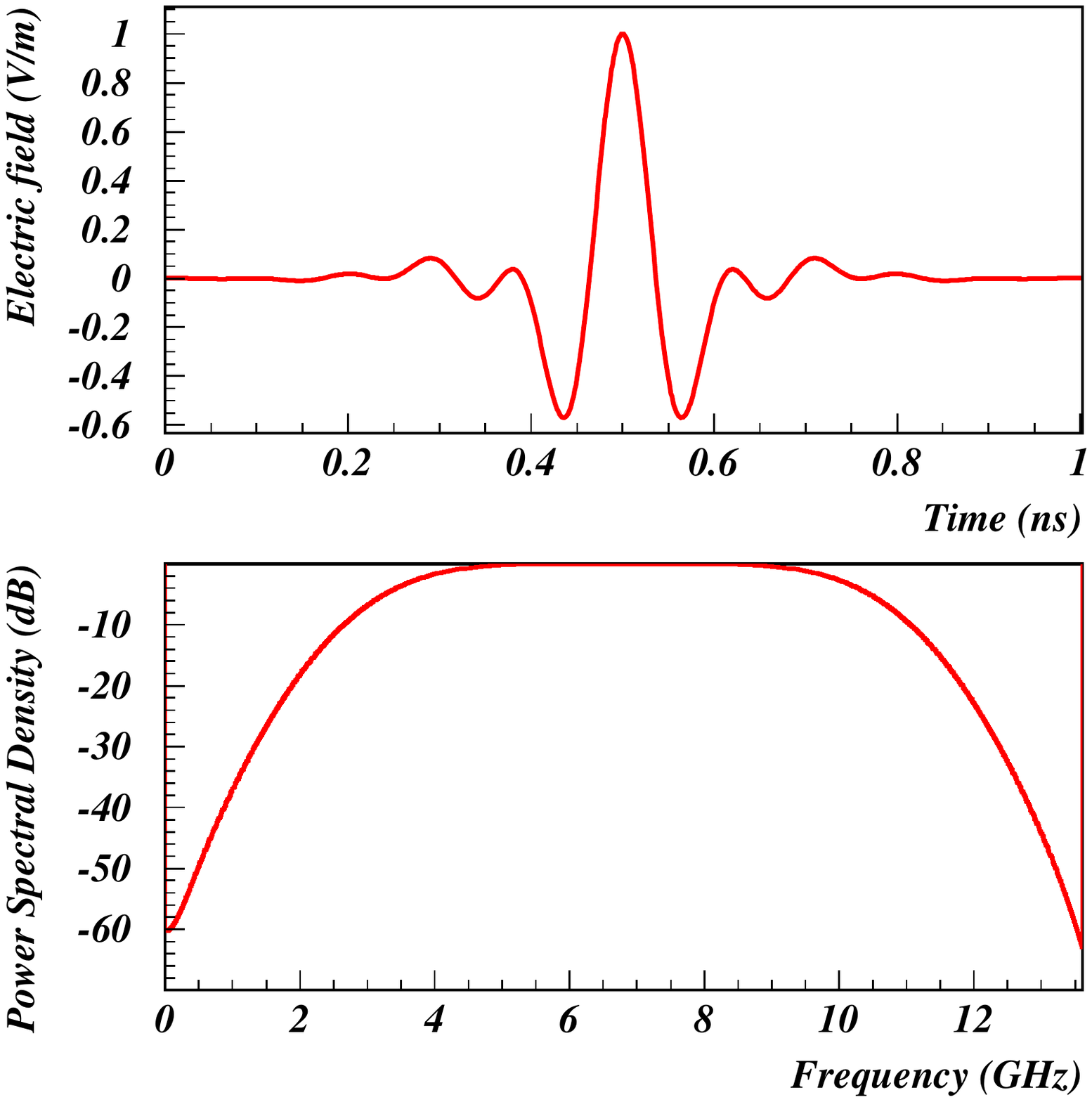}\hspace*{-1.7cm}
\includegraphics[scale=0.33]{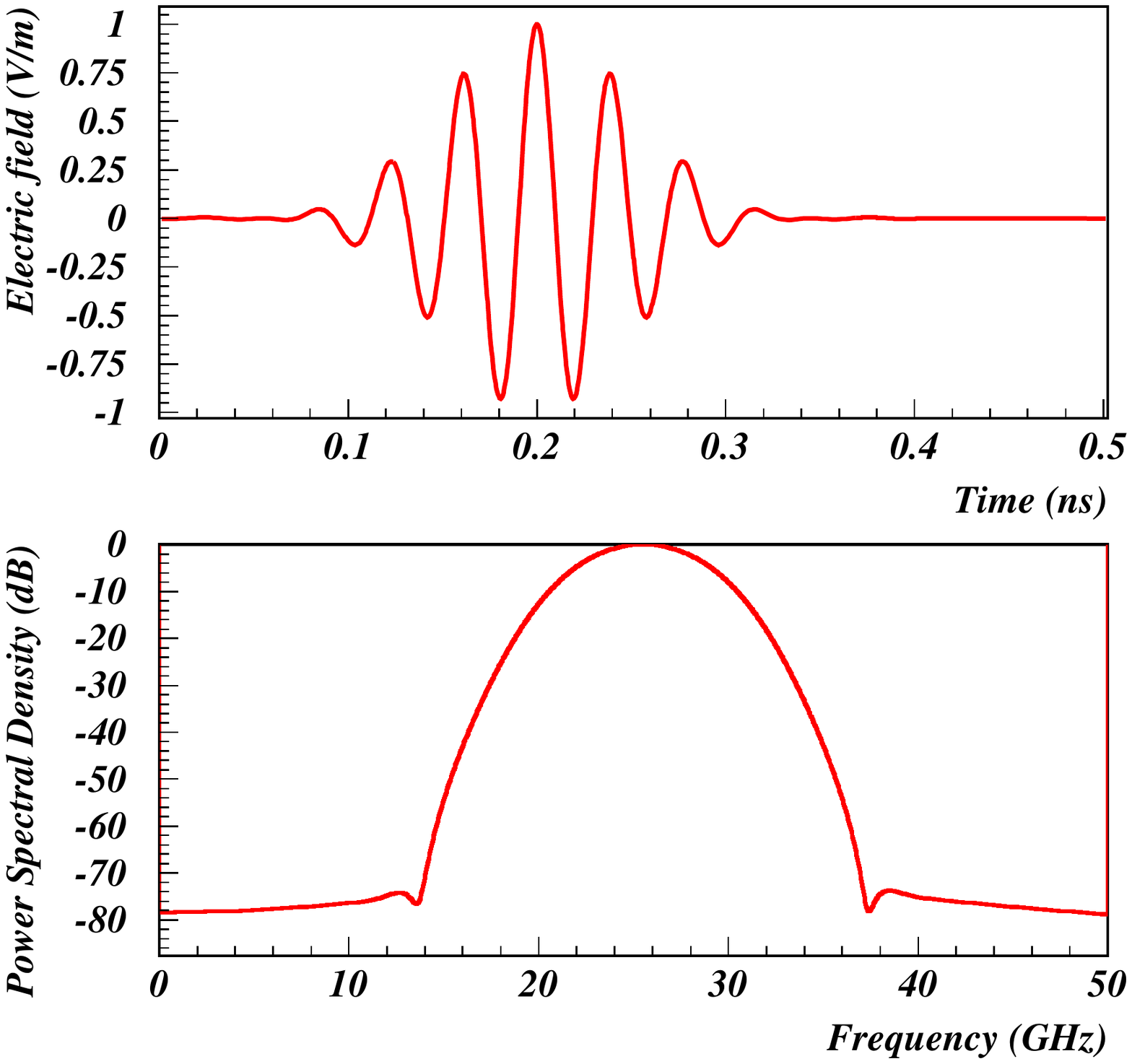}\hspace*{-1.4cm}
\includegraphics[scale=0.33]{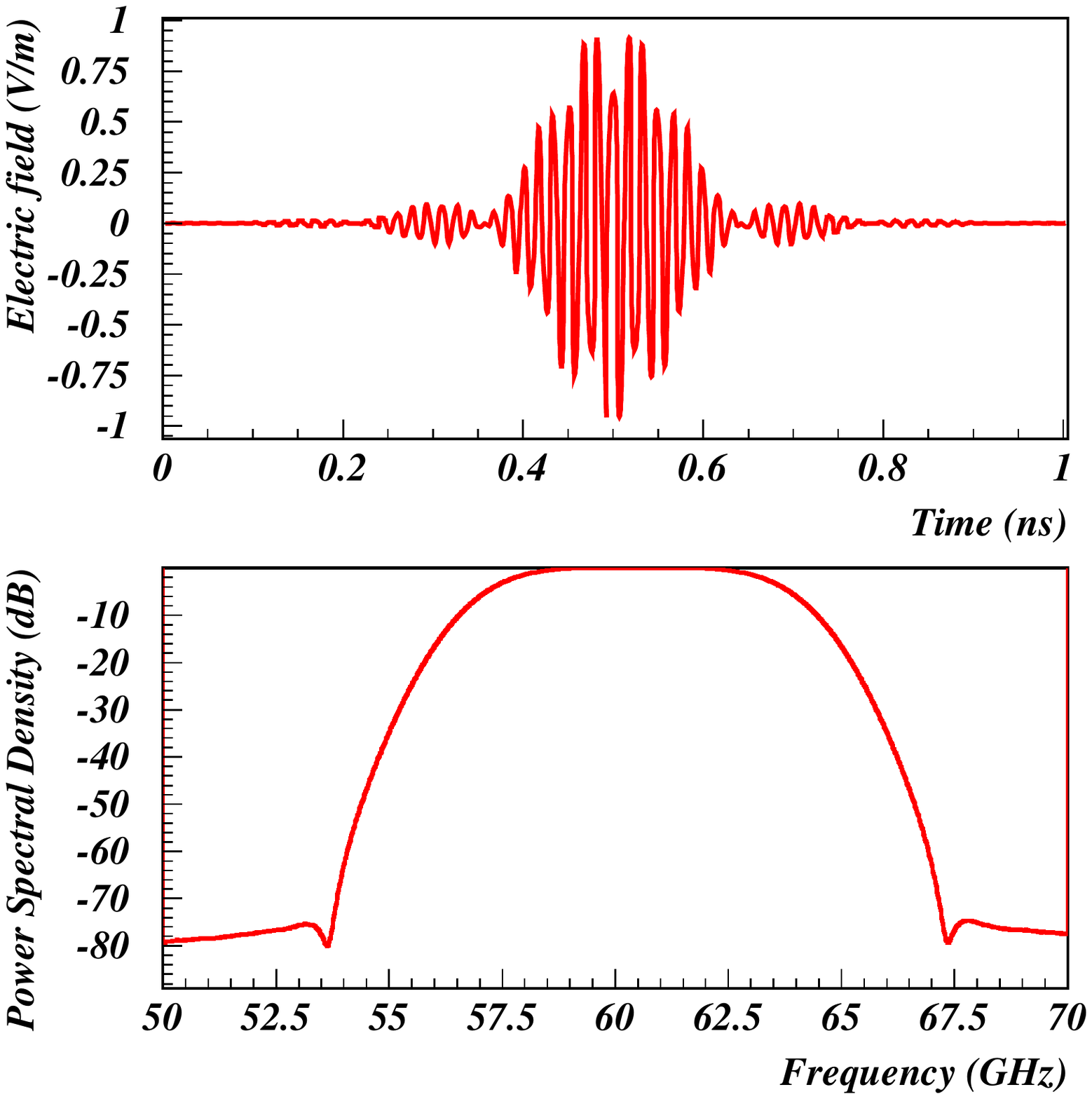}
\vspace*{- 2.5cm}
\end{center}
\caption{Shapes and frequency spectra of the UWB pulses described by Equation~\ref{pulse}. 
The amplitudes are normalized to $\rm 1 \; V/m$. 
From the left are the shape and the spectrum for the pulse in the
frequency region of $\rm 3.1 - 10.6 \; GHz$, pulse in the region of $\rm 22 - 29 \; GHz$,
and pulse in the region of $\rm 54 - 67 \; GHz$.}
\label{spectra}
\end{figure}

The eye was exposed to vertically polarized EM pulses, each covering 
one of three different frequency regions: $\rm 3.1 - 10.6 \; GHz$,  $\rm 22 - 29 \; GHz$, 
and $\rm 57 - 64 \; GHz$. The shape of the pulse is obtained by the 
inverse Fourier transformation  of the uniform spectral density 
of frequency width  $\rm \Delta f$ around the central frequency $\rm f_{c}$
of the frequency region. In the time domain, this transform is 
described by the function

\begin{equation} 
E=E_{0} {\sin[\pi  \Delta f (t-t_{0})] \over  {\pi \Delta f (t-t_{0})}}
\cos[2 \pi f_{c} (t-t_{0})] e^{ - {(t-t_{0})^{2}  \over  {2 \sigma^{2}}}}, 
\label{pulse}
\end{equation}
where $E_{0}$ is the pulse amplitude,  $t_{0}$ is the time shift of the pulse, and 
$\rm \sigma$ is the pulse width. The exponential term
 $\rm e^{ - {(t-t_{0})^{2}  \over  {2 \sigma^{2}}}} $ allows for a smooth rise and fall of the pulse
having, at the same time, a small effect on the pulse spectral density.
The numerical value of the parameter $t_{0}$ was the same for 
all the frequency regions, $t_{0}=3\sigma$. For
the frequency region of $\rm 3.1 - 10.6 \; GHz$:   $\rm\sigma = 166.7 \;  ps$, $\rm \Delta f = 7.5 \; GHz$ and  
$\rm f_{c}=6.85 \; GHz$, for the frequency region of $\rm 22 - 29 \; GHz$:  $\rm \sigma = 66.7 \; ps$, 
$\rm \Delta f = 7 \; GHz$ and  
$\rm f_{c}=25.5 \; GHz$; and for the frequency region of $\rm 57 - 64 \; GHz$:  $\rm \sigma = 66.7 \; ps$,
$\rm \Delta f = 7 \; GHz$ and  
$\rm f_{c}=60.5 \; GHz$. The shapes and the frequency 
spectra of the pulses are plotted in Figure~\ref{spectra}.

Understanding the propagation of a UWB pulse is more demanding 
than understanding the propagation of a continuous EM wave. 
For some pulse frequencies the exposed 
sample looks like a conductor while for others it looks like a dielectric. The amount of 
the conduction currents compared to the displacement currents changes 
over the frequency spectrum of the EM pulse. In this paper we have restricted 
ourselves to studying the propagation of the 
pulses with almost uniform power spectra in the frequency range imposed by the 
FCC rules.

\section{Dielectric Properties of the Exposed Eye Tissue}

Electromagnetic properties of a tissue are normally expressed in terms
of frequency-dependent dielectric properties and conductivity.
When dealing with an UWB electromagnetic pulse, it is crucial
that these properties are properly described over a large frequency range. 
One of the most accepted models describing the dielectric properties of 
a tissue is the  Cole-Cole model. The references  
(Gabriel 1996, Gabriel C {\it et al} 1996, Gabriel S {\it et al} 1996))
provide the four-term Cole-Cole parametrization of 
the dielectric properties of the ocular tissue used in the
present work

\begin{equation}
\varepsilon(\omega) = \varepsilon_{\infty} + \sum_{k=1}^{4}
{\Delta \varepsilon_{k} \over {1+(i\omega\tau_{k})^{1-\alpha_{k}}}} 
+ {\sigma  \over i\omega \varepsilon_{0}}.
\label{CC}
\end{equation}
In this equation $i=\sqrt{-1}$, $\varepsilon_{\infty}$ is the permittivity in the 
terahertz frequency range, $\Delta \varepsilon_{k}$ are the changes in 
permittivity in a specified frequency range, $\tau_{k}$ are the relaxation 
times, $\sigma$ is the ionic conductivity, and  $\alpha_{k}$ are the coefficients
specific for the Cole-Cole model. They constitute up to 14 real parameters of a 
fitting procedure. 

Application of the Cole-Cole model is problematic for FDTD. It requires time 
consuming numerical integration techniques and makes computation unacceptably slow.
If, instead of a Cole-Cole parametrization, the Debye model 

\begin{equation}
\varepsilon(\omega) = \varepsilon_{\infty} + \sum_{k=1}^{N}
{ \Delta \varepsilon_{k} \over {1+i\omega\tau_{k}}}+ {\sigma  \over i\omega \varepsilon_{0}}
\label{Dpar}
\end{equation}
is used, the computation time can be reduced by an order of magnitude using very 
efficient piecewise-linear recursive convolution (PLRC) method (Luebbers {\it et al} 1990, 
1991, Luebbers and Hunsberger 1992, Kunz and Luebbers 1993, Taflove and Hagness 2000). 
But, it is also important that the Debye model 
provides an equally accurate description of the physical properties of a biological tissue.  

\begin{figure}
\begin{center}
\includegraphics[scale=0.38]{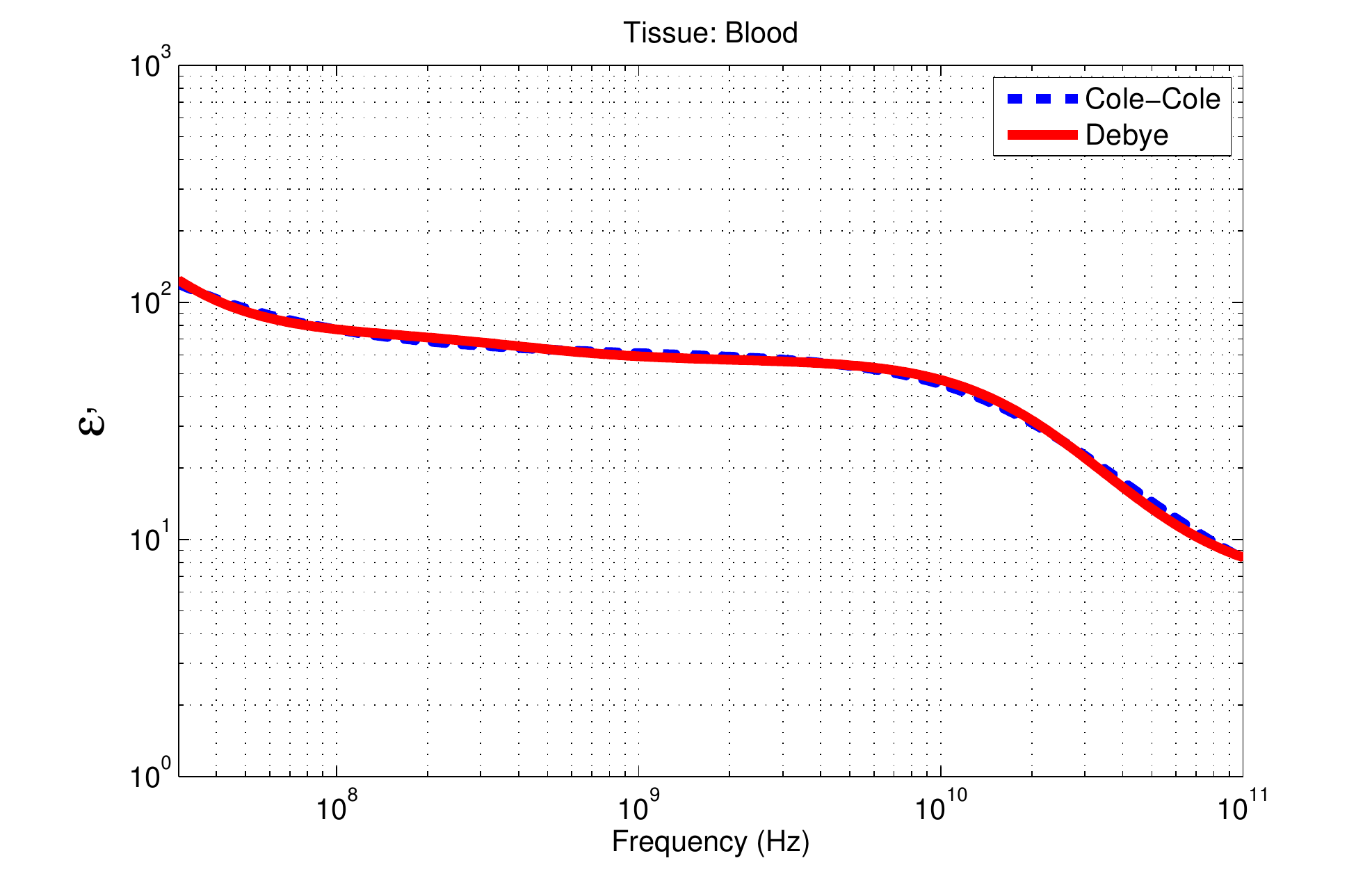}
\includegraphics[scale=0.38]{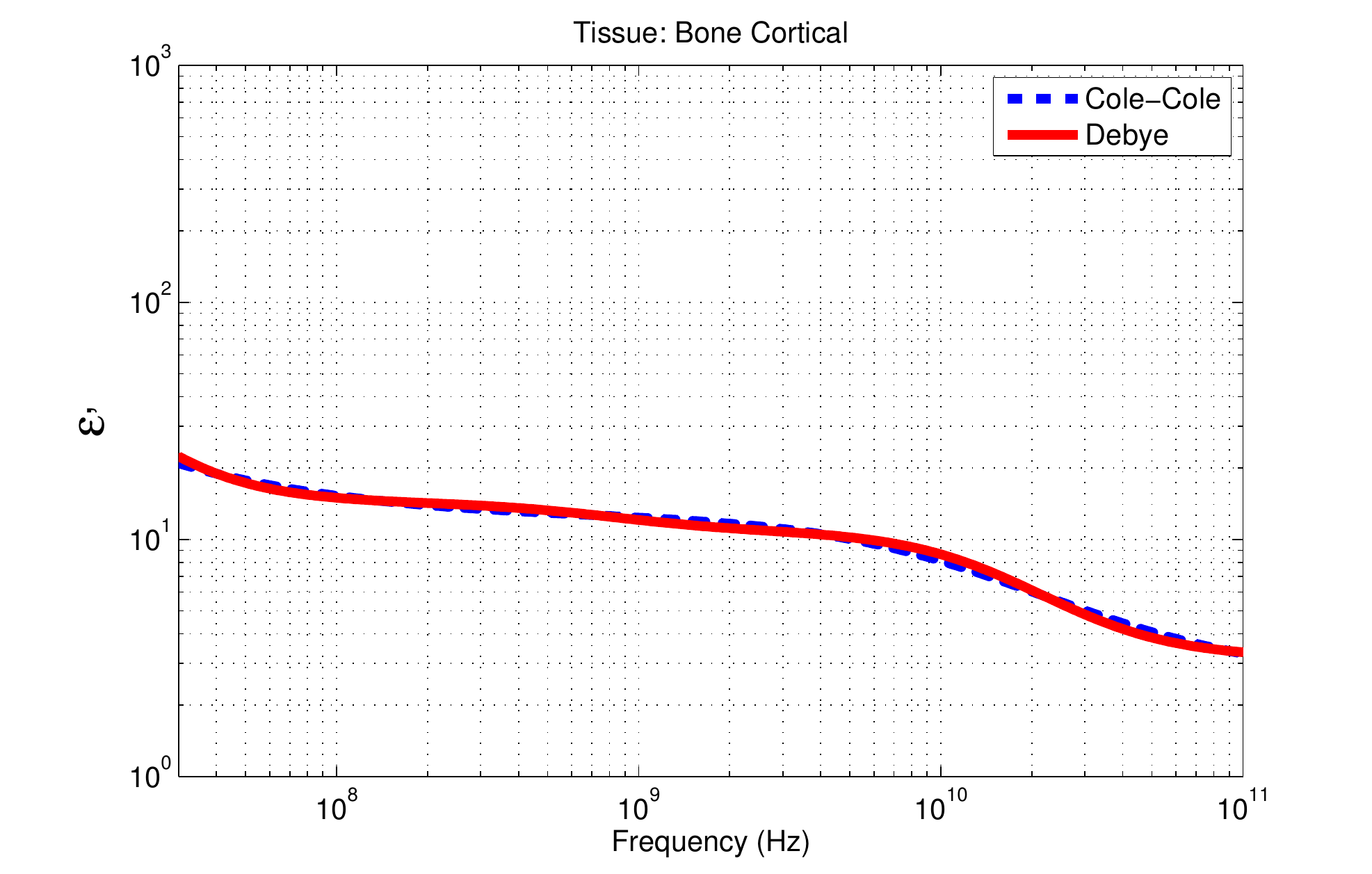}
\includegraphics[scale=0.38]{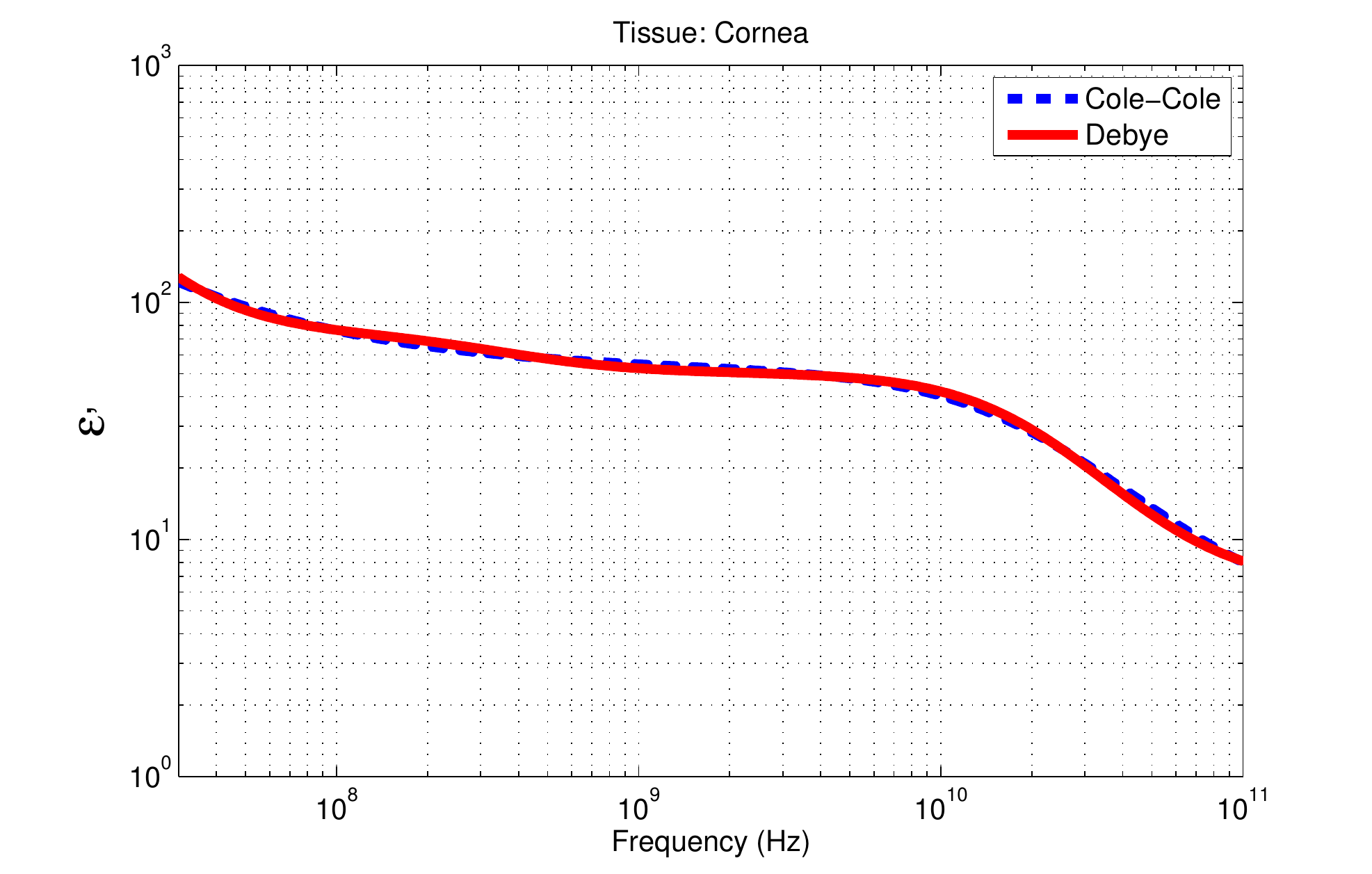}
\includegraphics[scale=0.38]{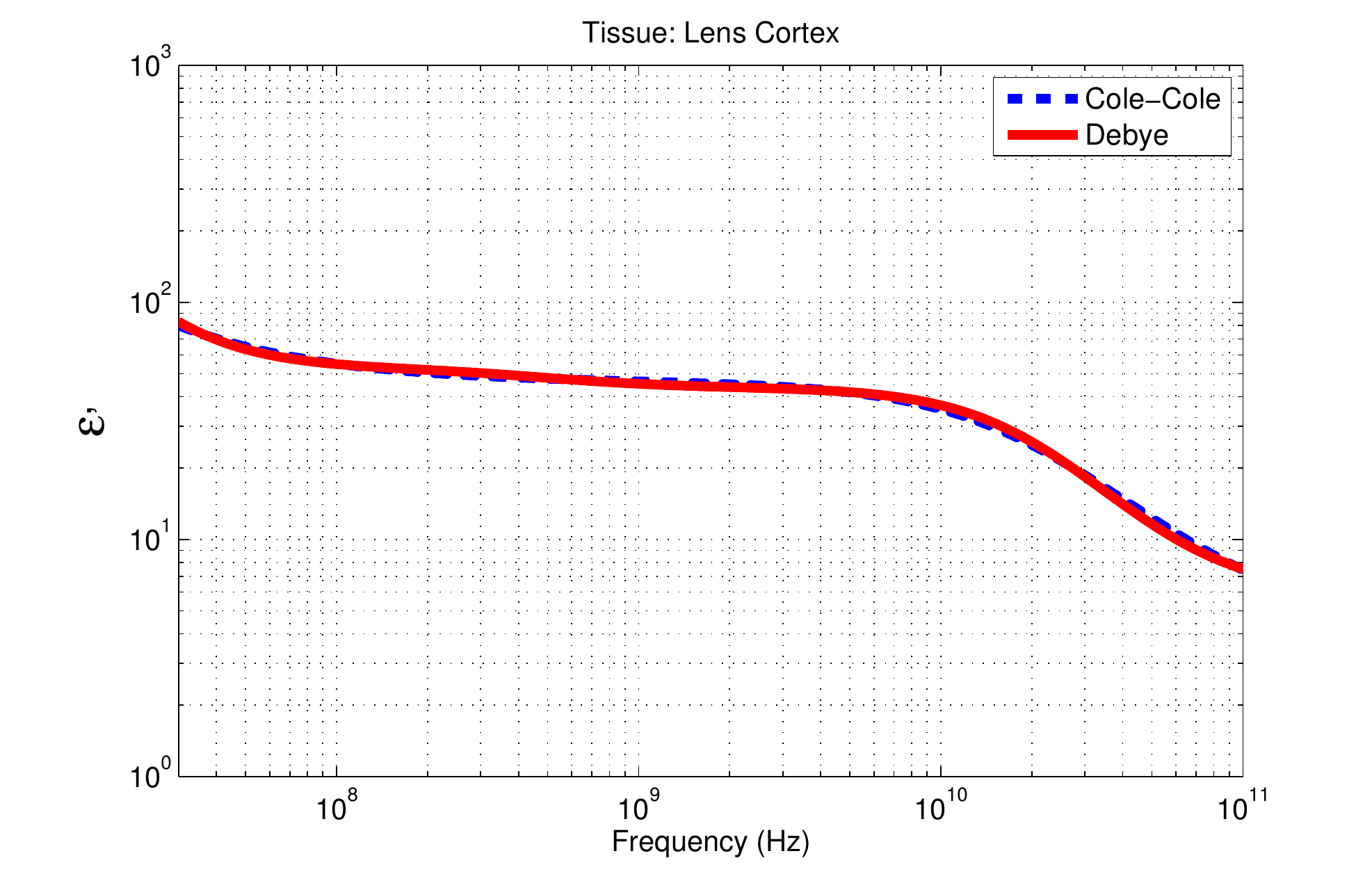}
\includegraphics[scale=0.38]{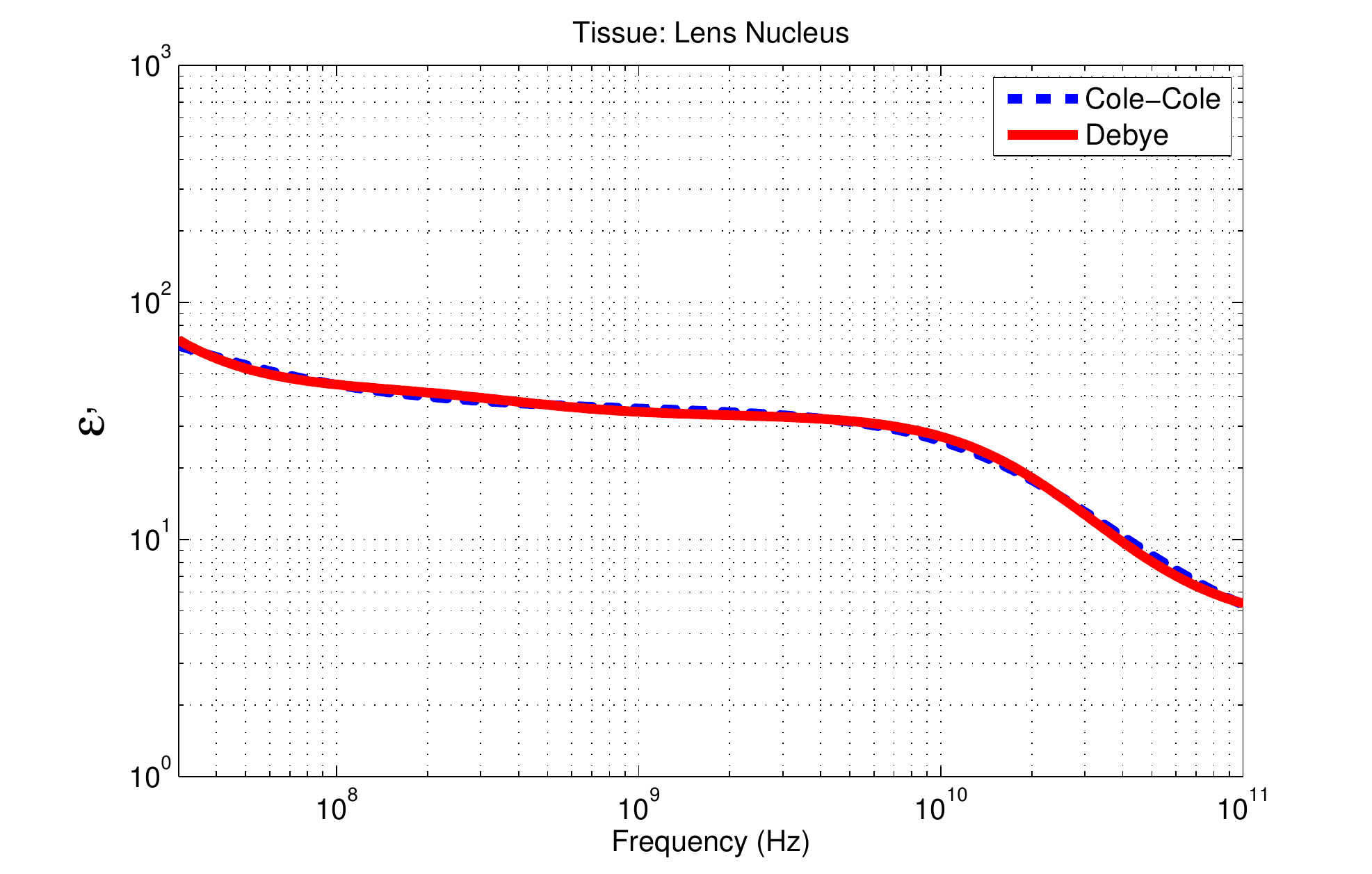}
\includegraphics[scale=0.38]{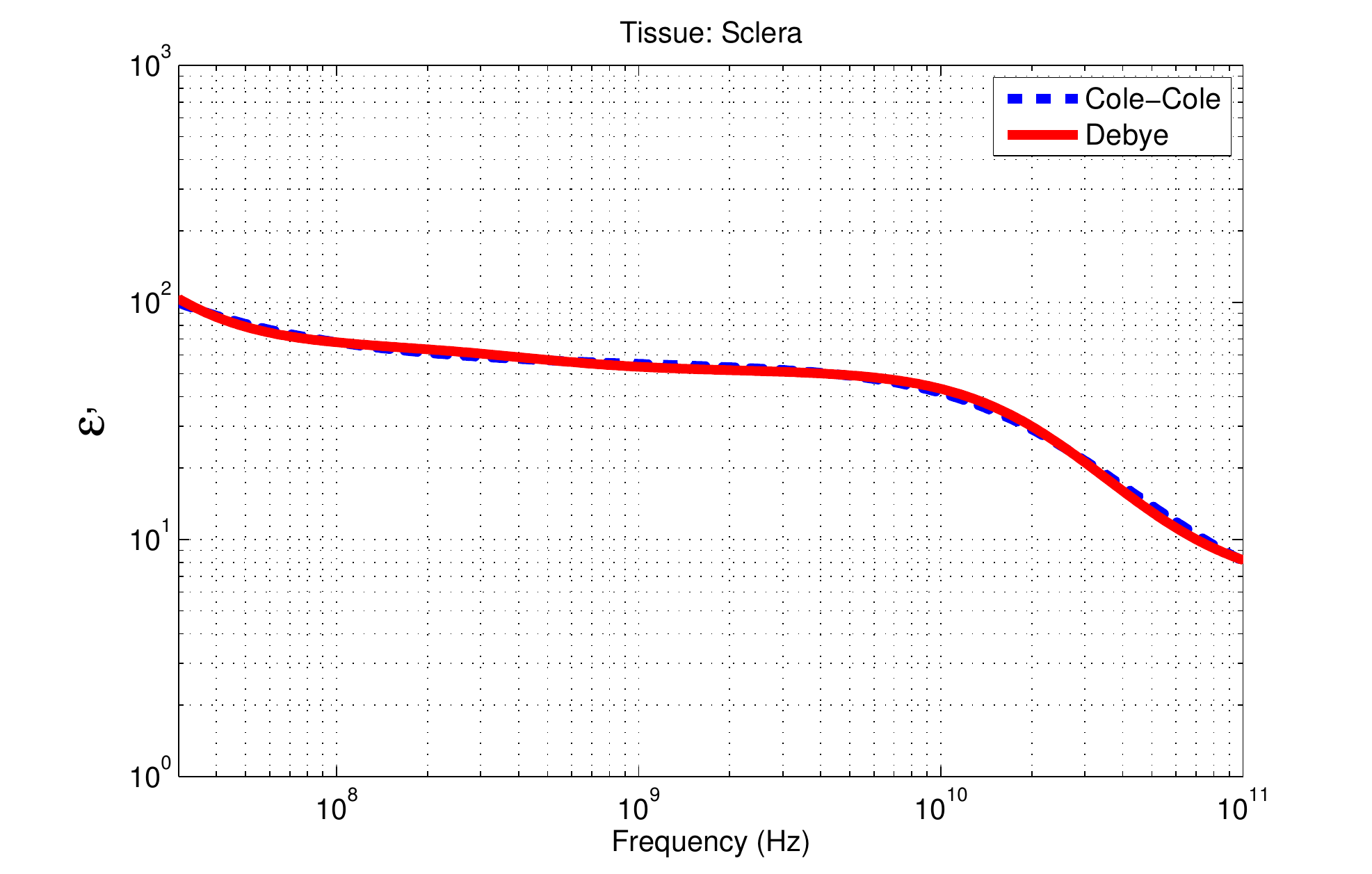}
\includegraphics[scale=0.38]{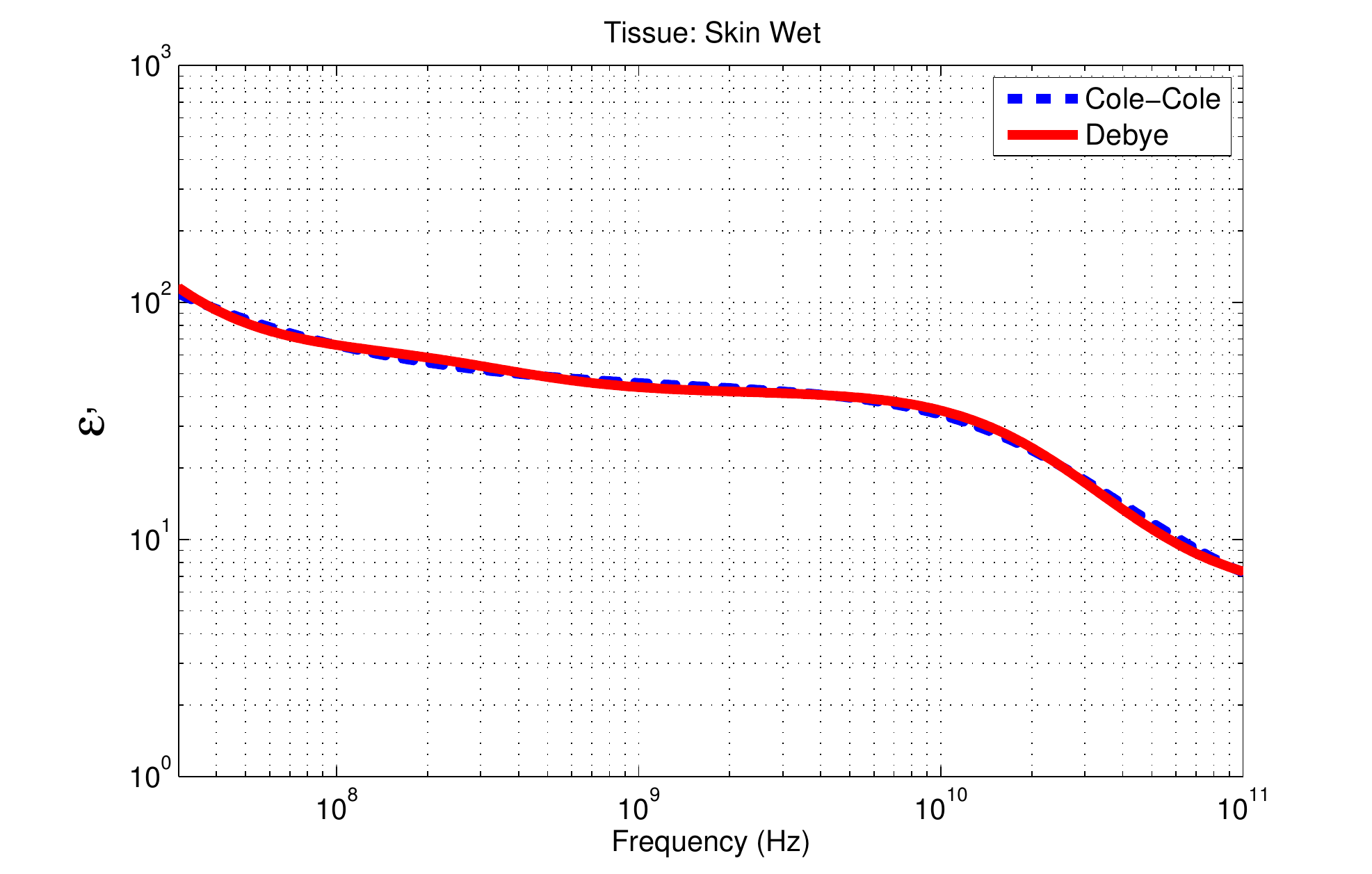}
\includegraphics[scale=0.38]{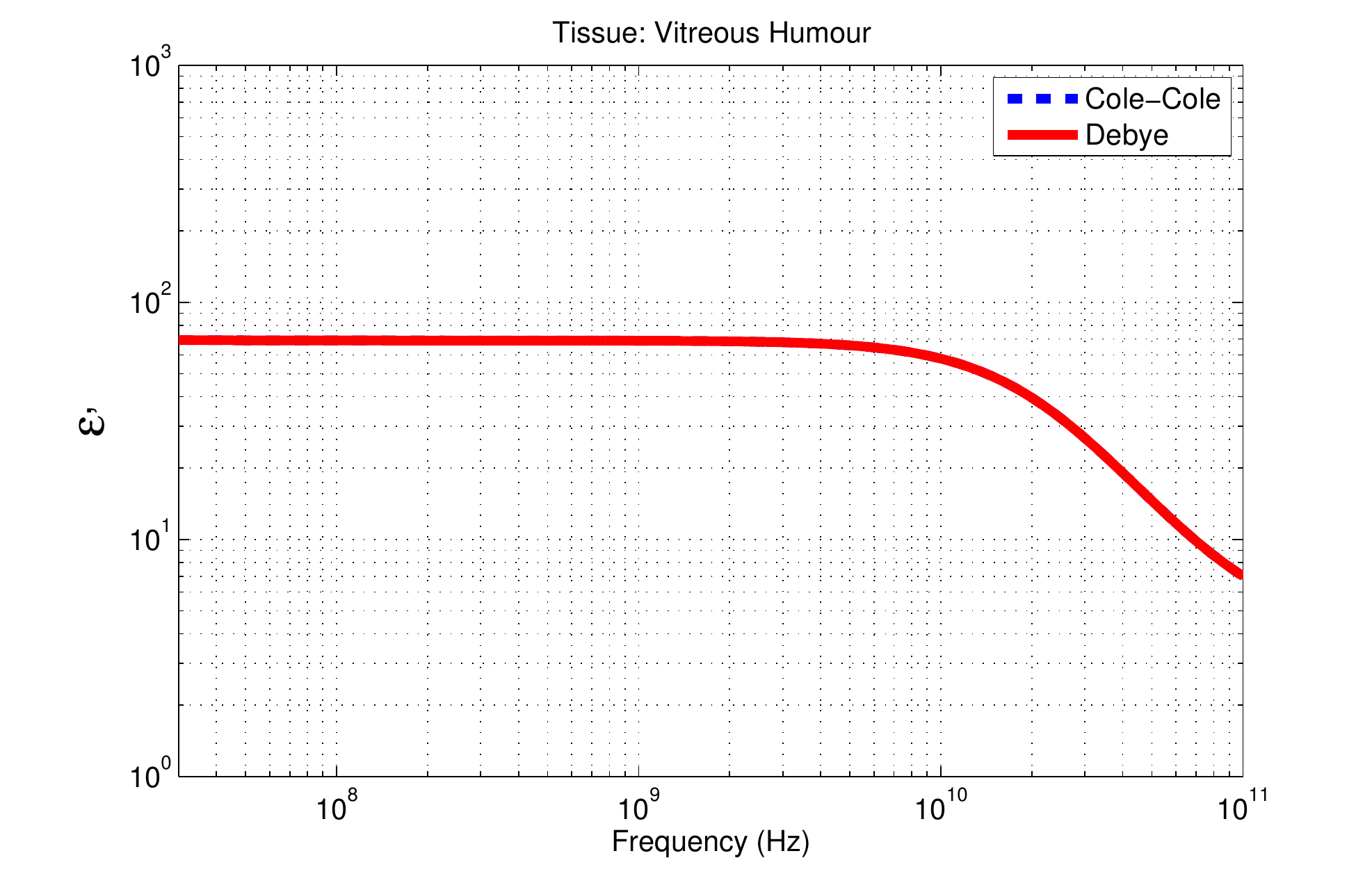}
\end{center}
\caption{Comparison of the relative permittivity of the tissues used in the computation 
parametrized by the Debye model (Equation \ref{Dpar2}; red solid line)
and the Cole-Cole model (Equation \ref{CC}; blue dashed line). Parametrization covers 
frequency range from 30 MHz to 100 GHz.} 
\label{dielectrics}
\end{figure}

In his thesis, G. R. Lugo (Lugo 2006) used an accurate, robust and efficient vector fitting technique
(VECTFIT) (Gustavsen and Semlyen  1999) to replace the Cole-Cole parametrization by a multi-term
Debye parametrization with no loss in the precision. His procedure is used in this paper.
While the accuracy of the Debye parameterization increases with the number of terms, 
so does the computational time and the requirement on the memory, and an optimal number
of terms has to be selected. It was found that in the frequency
region considered in this paper the accuracy of the three-term Debye model 
\begin{equation}
\varepsilon(\omega) = \varepsilon_{\infty} 
+{{\varepsilon_{s1}-\varepsilon_{\infty}} \over {1+i\omega\tau_{1}}}
+{{\varepsilon_{s2}-\varepsilon_{\infty}} \over {1+i\omega\tau_{2}}}
+{{\varepsilon_{s3}-\varepsilon_{\infty}} \over {1+i\omega\tau_{3}}}
\label{Dpar2}
\end{equation}
is approximately the same as the accuracy of the corresponding Cole-Cole model.
The parameters of the three-term Debye model for the tissue used in the 
calculation are presented in Table \ref{tab1} and the comparison between the two 
models is shown in Figure \ref{dielectrics}.

\begin{table}
\begin{center}
\begin{tabular}{lcccccccc}
\multicolumn{1}{c}{Material} {\vline}&
\multicolumn{1}{c}{$\varepsilon_{\infty}$} {\vline}&
\multicolumn{1}{c}{$\varepsilon_{s1}$} {\vline}&
\multicolumn{1}{c}{$\varepsilon_{s2}$} {\vline}&
\multicolumn{1}{c}{$\varepsilon_{s3}$} {\vline}&
\multicolumn{1}{c}{$\tau_{1} (s)$} {\vline}&
\multicolumn{1}{c}{$\tau_{2} (s)$} {\vline}&
\multicolumn{1}{c}{$\tau_{3} (s)$} {\vline}&
\multicolumn{1}{c}{$\sigma (S/m)$} {\vline}\\
\hline\hline
   Blood                        &  6.5  & 50.7& 16.2& 9835. &  $7.95 \; 10^{-12}$  & $4.08 \; 10^{-10}$  & $7.35 \; 10^{-8}$ & 0.7\\
   Bone  Cortical        &  3.2  & 7.7  & 3.4  &  137.   &  $1.01 \; 10^{-11}$  & $2.07 \; 10^{-10}$  & $2.11 \; 10^{-8}$ & 0.02\\
   Cornea                     &  6.3  & 44.2 & 22.9 & 6122. &  $7.73 \; 10^{-12}$  & $4.75 \; 10^{-10}$  &  $5.59 \; 10^{-8}$& 0.4\\
   Lens Cortex            &  5.9  & 37.9 & 8.7 & 3873. &  $7.57 \; 10^{-12}$  & $3.33 \; 10^{-10}$  &  $5.96 \; 10^{-8}$ & 0.3\\
   Lens Nucleus         &  4.3  & 29.1  & 9.9 &  996. &  $8.34  \; 10^{-12}$  & $4.33 \; 10^{-10}$  & $3.24 \; 10^{-8}$ & 0.2\\
   Muscle                     &  6.5  & 45.2  & 11.9 & 5018. &  $7.00 \; 10^{-12}$  & $3.71 \; 10^{-10}$  & $6.60 \; 10^{-8}$  & 0.2\\
   Sclera                       &  6.3  & 45.3  & 13.5 & 6912. &  $7.61 \; 10^{-12}$  & $3.86 \; 10^{-10}$  &  $7.07 \; 10^{-8}$ & 0.5\\
   Skin  Wet                  & 5.9   & 36.0 & 21.1 &  1417. &  $7.75 \; 10^{-12}$  & $4.85 \; 10^{-10}$  &  $2.70 \; 10^{-8}$ & 0.0\\
   Vitreous Humour   &  4.2  & 65.1 &  $3.6 \; 10^{7}$ & - &  $7.32 \; 10^{-12}$  & $2.14 \; 10^{-4}$  &  - & 1.5\\

\end{tabular}
\end{center}
\caption{Debye parameters for the tissues used in the computation. Static conductivity $\sigma$
is from Reference (Gabriel C {\it et al} 1996).}
\label{tab1}
\end{table}

\section{Field Calculation and Data Representation}

The quantities required to understanding the mechanisms and the consequences of 
the interaction between the electromagnetic radiation and biological tissue are obtained 
from the the values of electric and magnetic field components inside the tissue.  
Those components are accurate on the level of our knowledge of the 
physical properties of the exposed material.
While the rapid developments in computer technology allow for the improvement of 
the geometrical description and for a longer exposure time calculation, for a reliable output, 
the knowledge of the physical properties of the material must be complete and correct. 
This is particularly
important when simulating the interaction of UWB radiation with biological material.

\begin{figure}
\begin{center}
\includegraphics[scale=0.44]{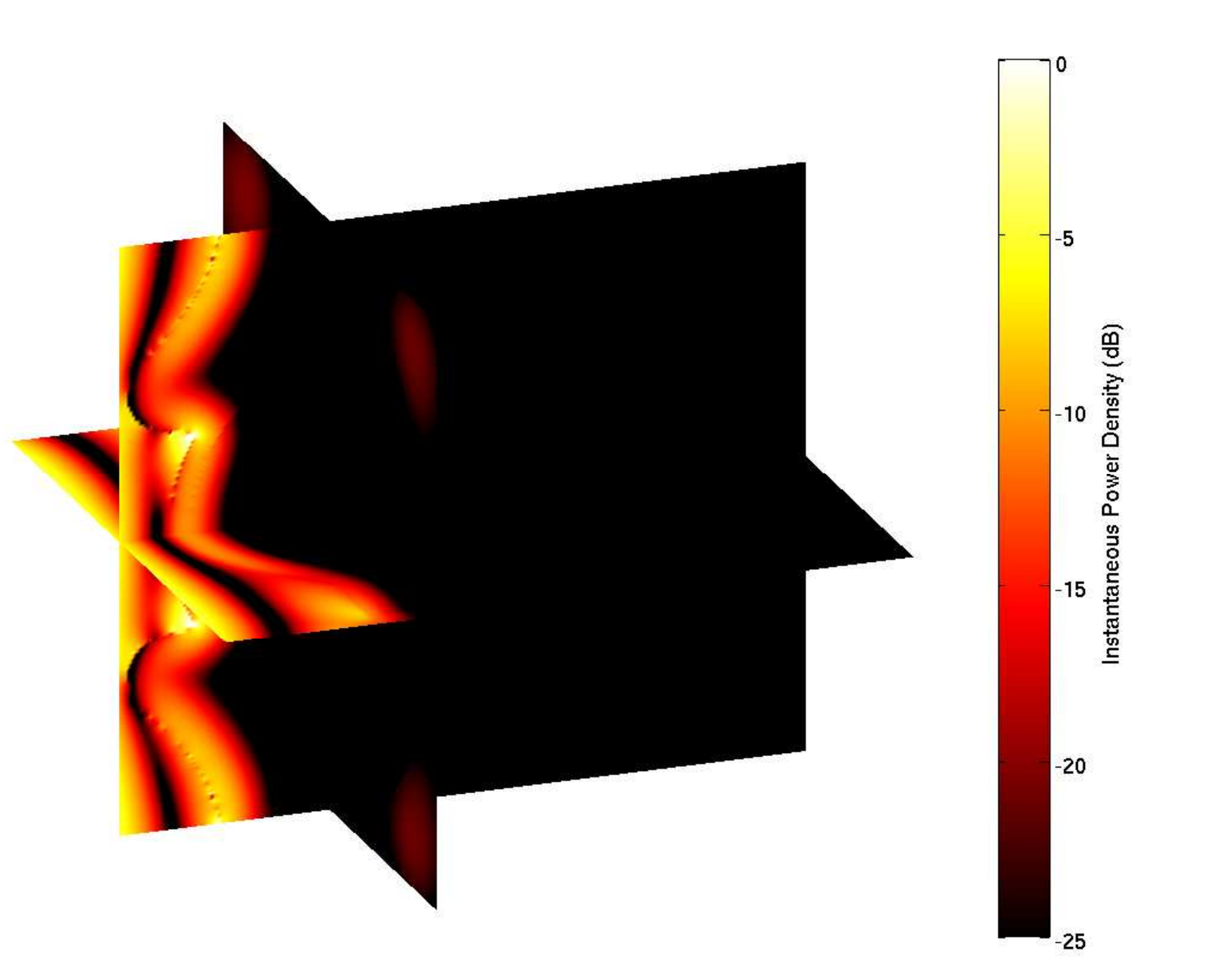}
\includegraphics[scale=0.44]{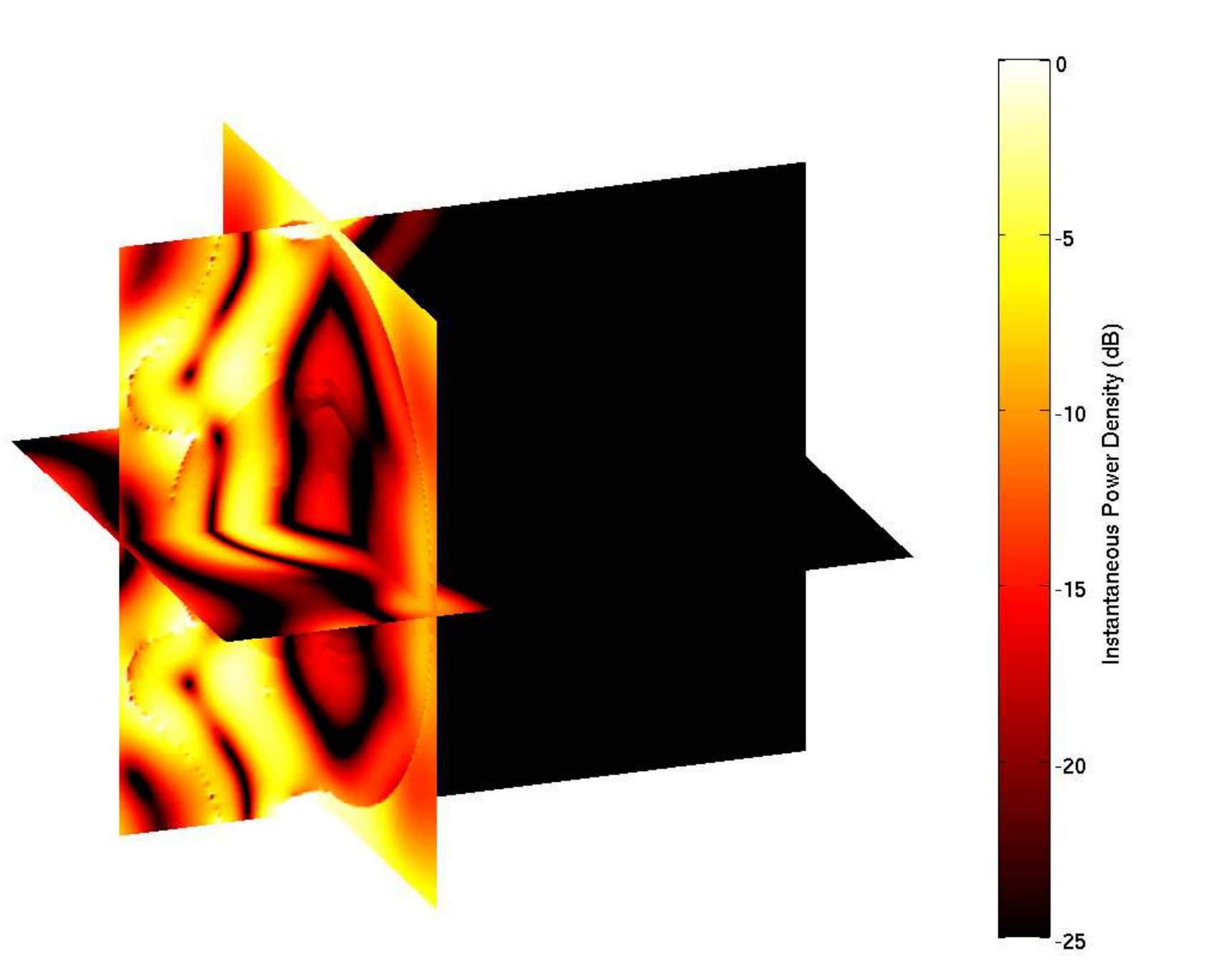}
\includegraphics[scale=0.44]{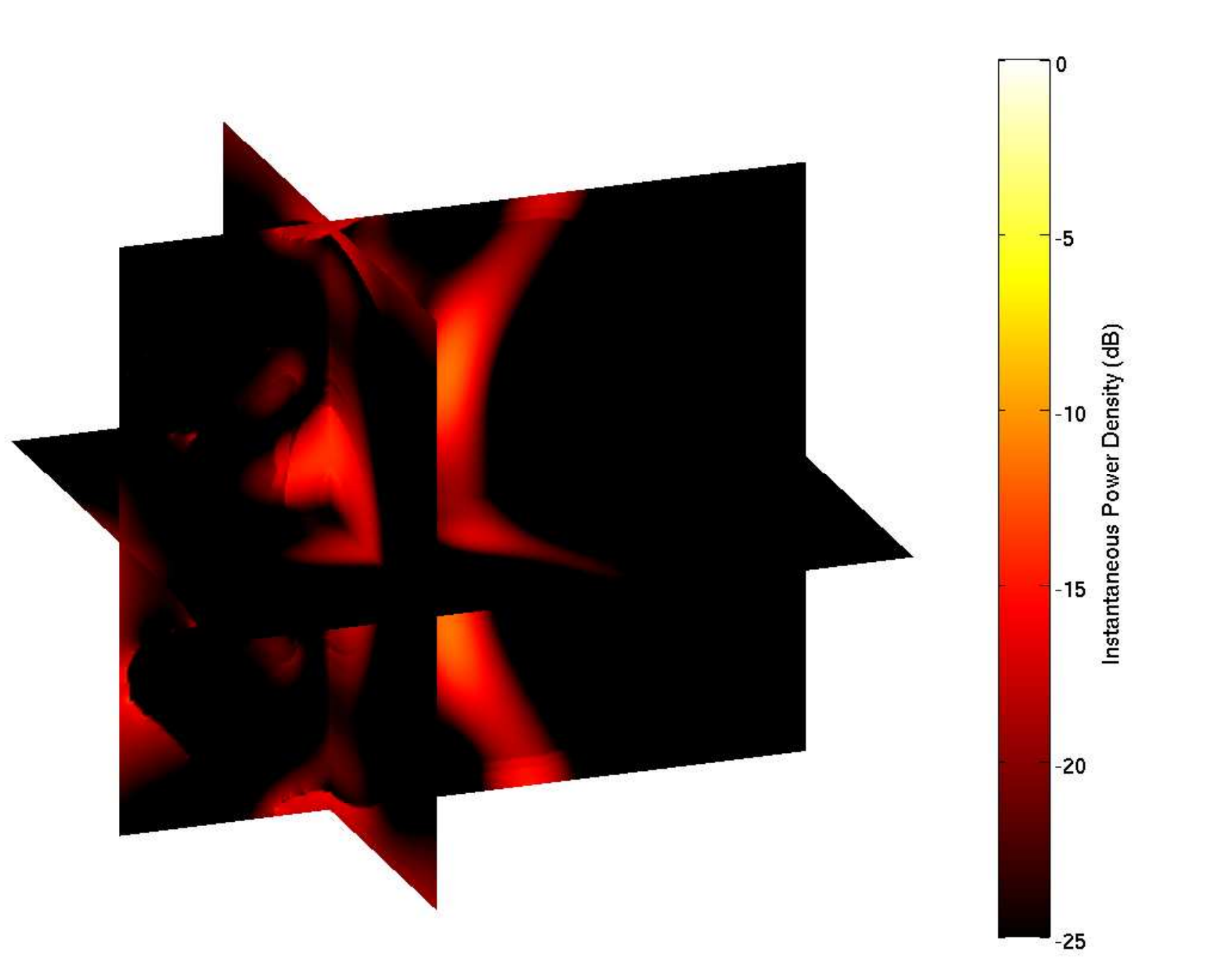}
\includegraphics[scale=0.44]{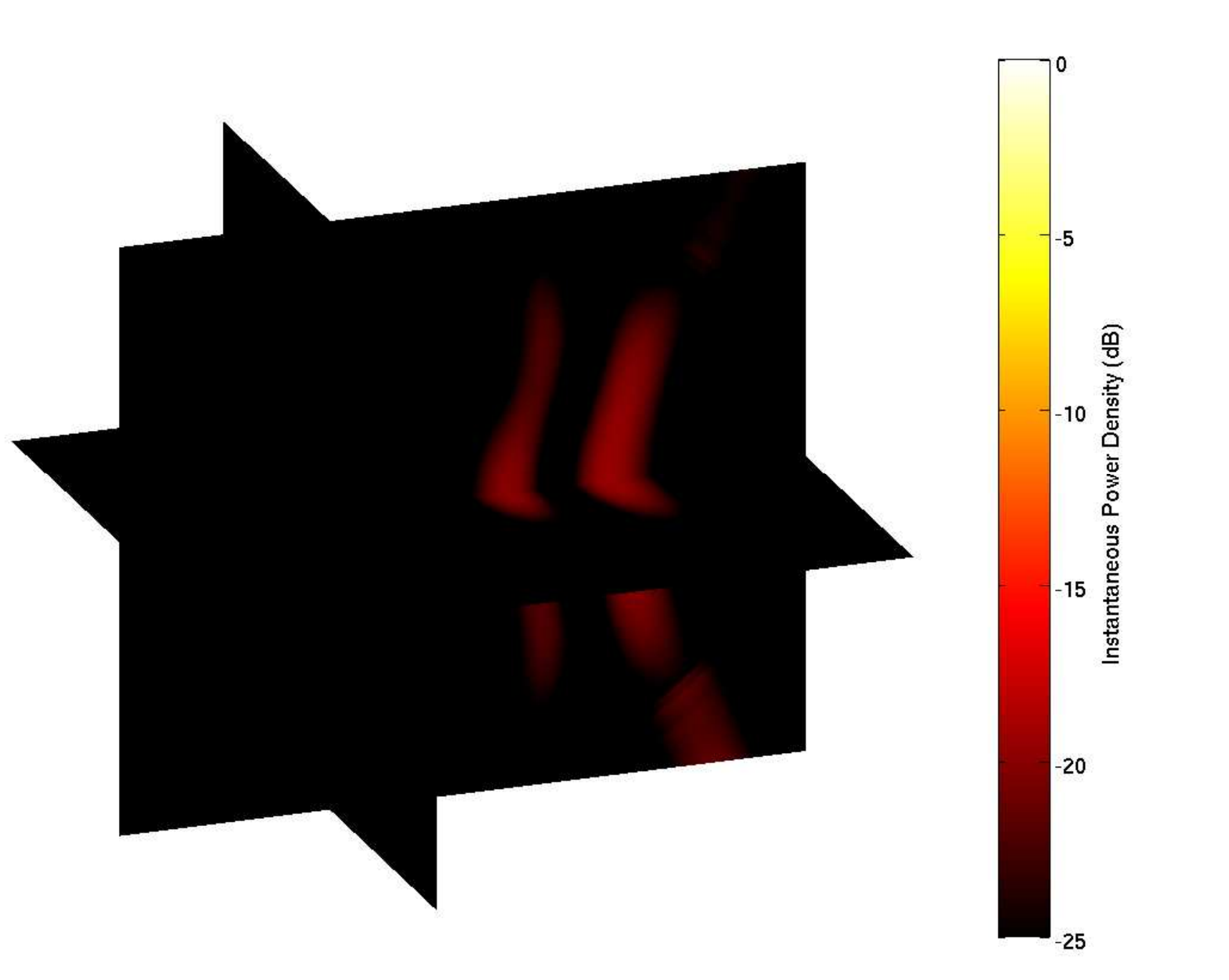}
\end{center}
\caption{Four different stages of penetration of EM pulse, in the
frequency region of $\rm 3.1 - 10.6 \; GHz$,  into the eye. The contours 
represent the instantaneous power density in decibels. Brighter regions correspond to 
higher power density.} 
\label{snaps}
\end{figure}

\begin{figure}
\begin{center}
\includegraphics[scale=0.44]{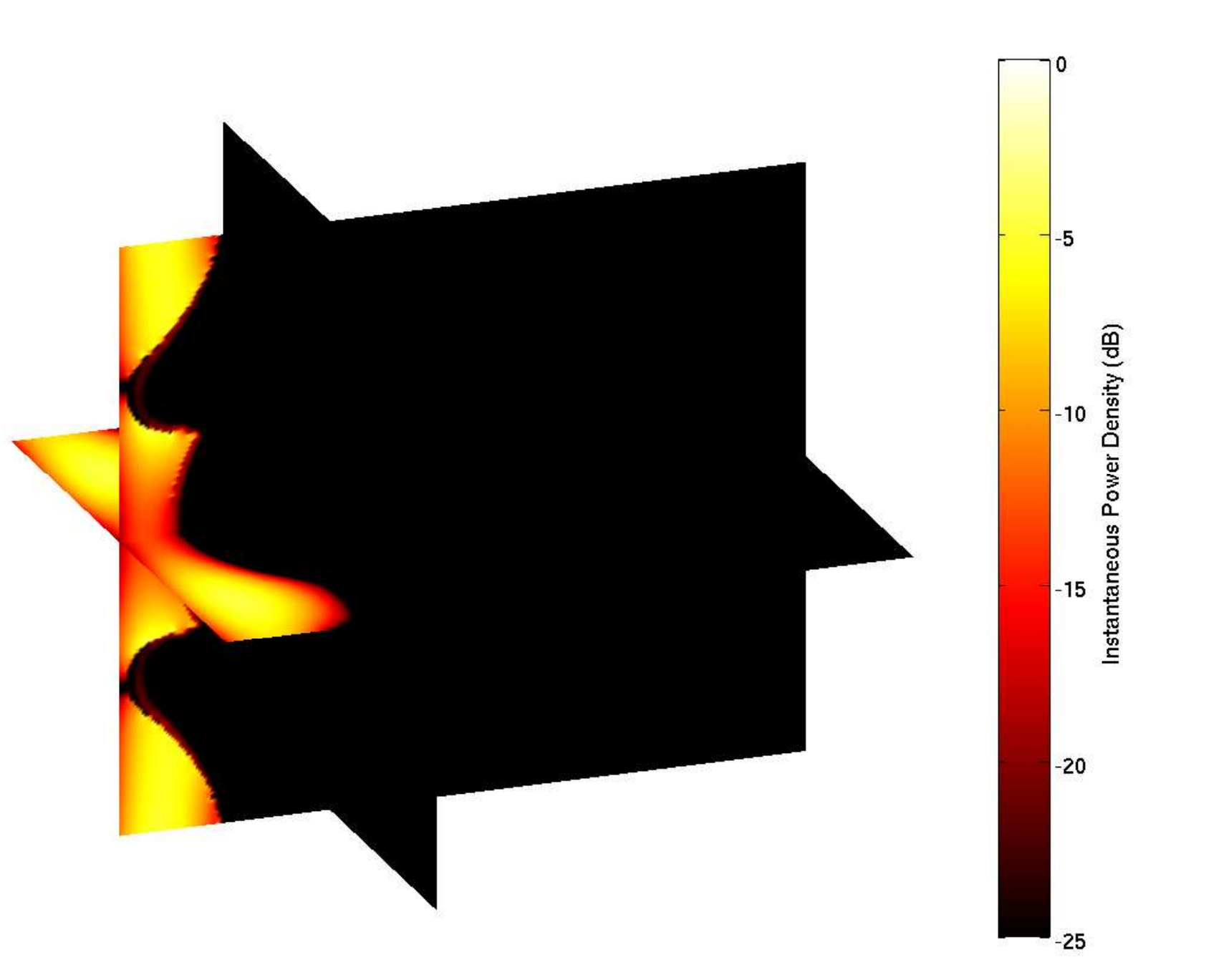}
\includegraphics[scale=0.44]{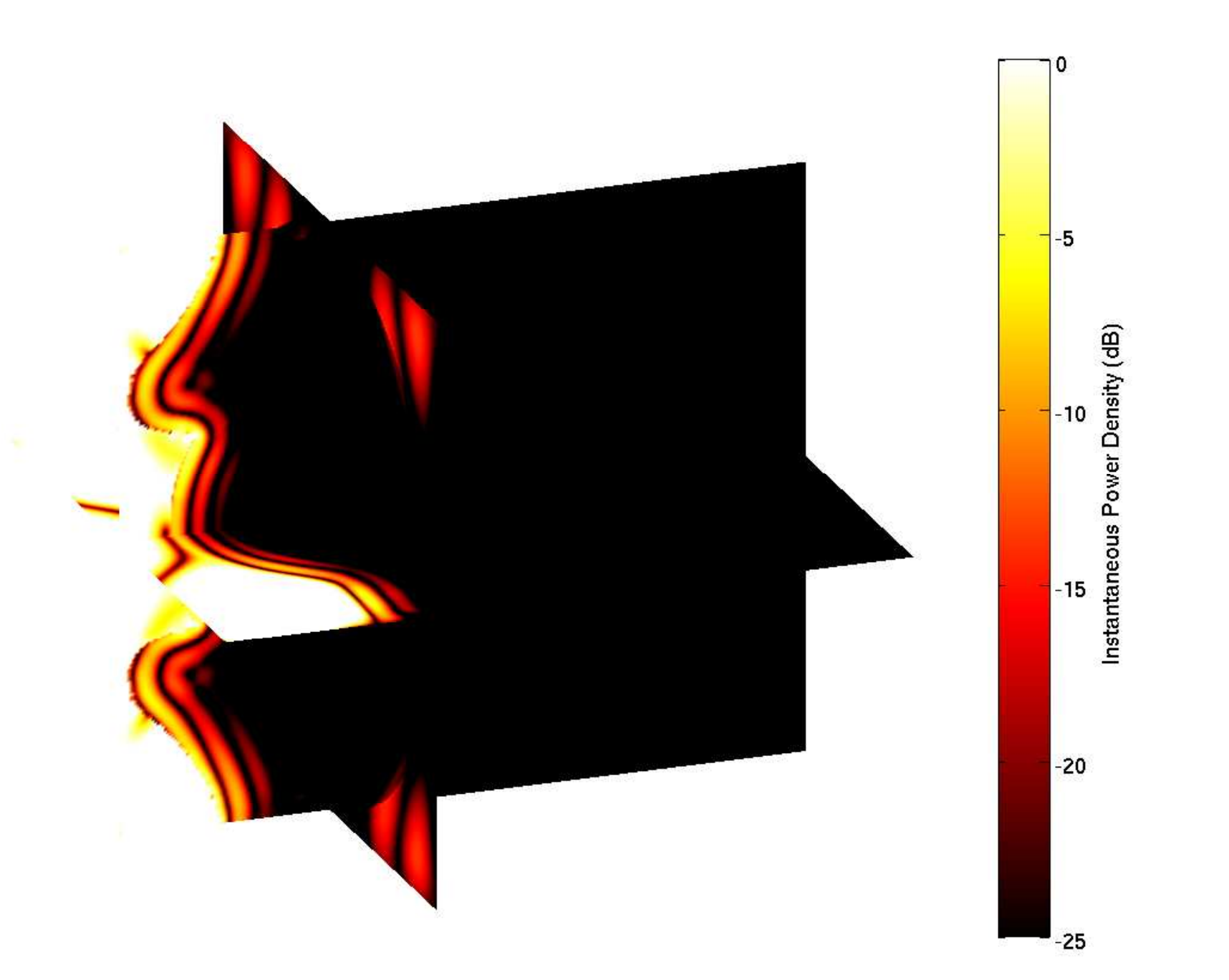}
\includegraphics[scale=0.44]{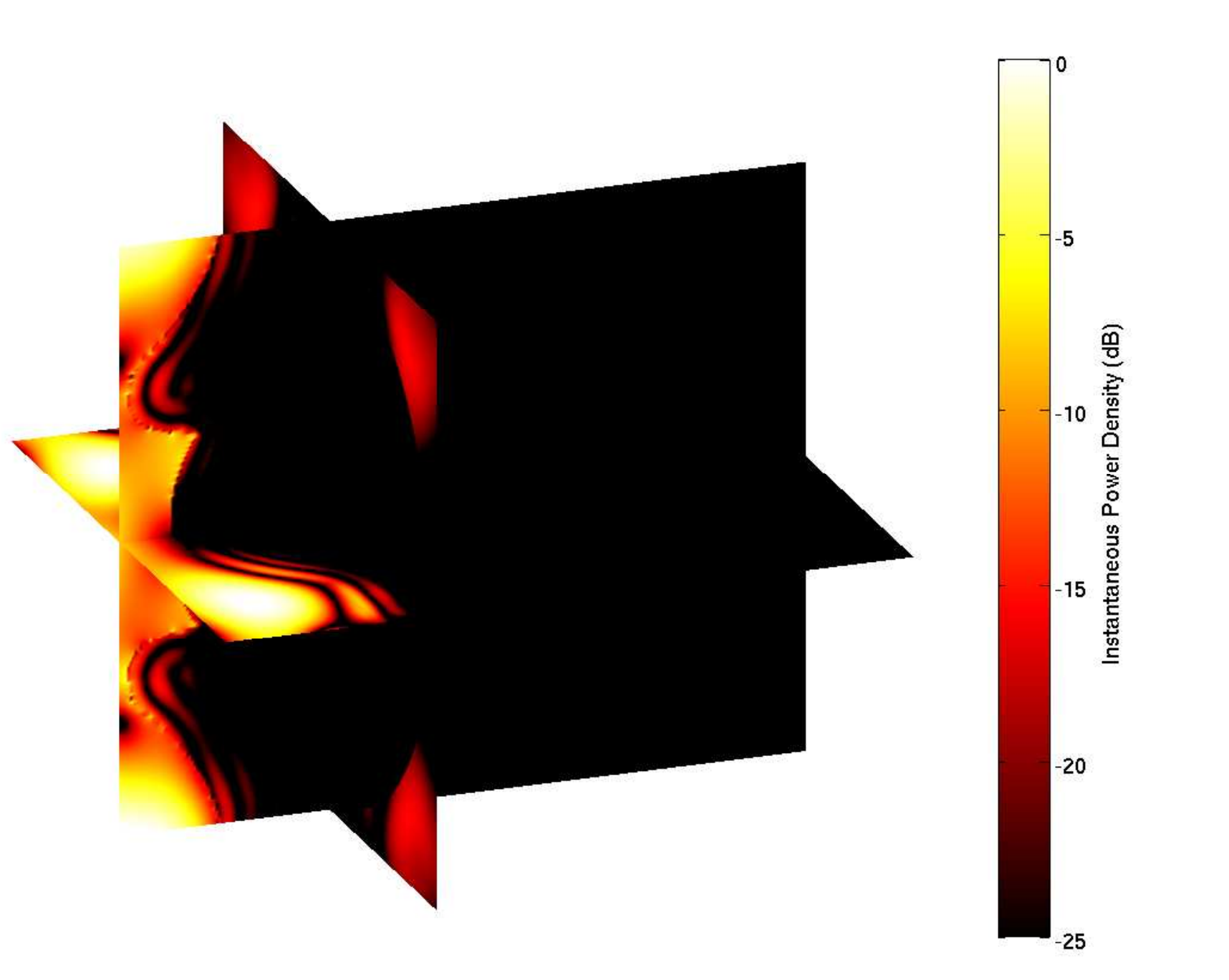}
\includegraphics[scale=0.44]{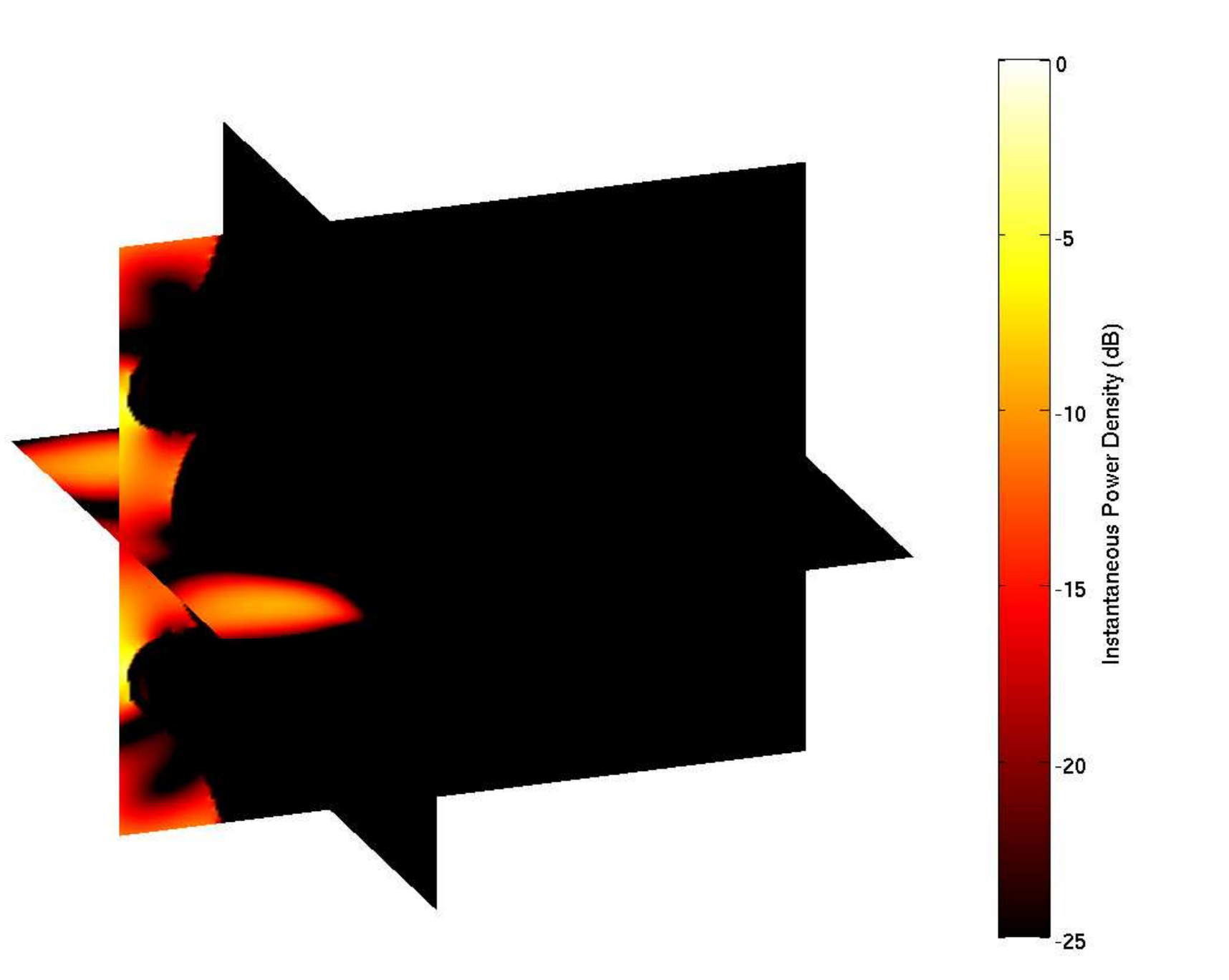}
\end{center}
\caption{Four different stages of penetration of EM pulse, in the
frequency region of $\rm 22 - 29 \; GHz$,  into the eye. The contours 
represent the instantaneous power density in decibels. Brighter regions correspond to 
higher power density.} 
\label{snaps2}
\end{figure}

\begin{figure}
\begin{center}
\includegraphics[scale=0.44]{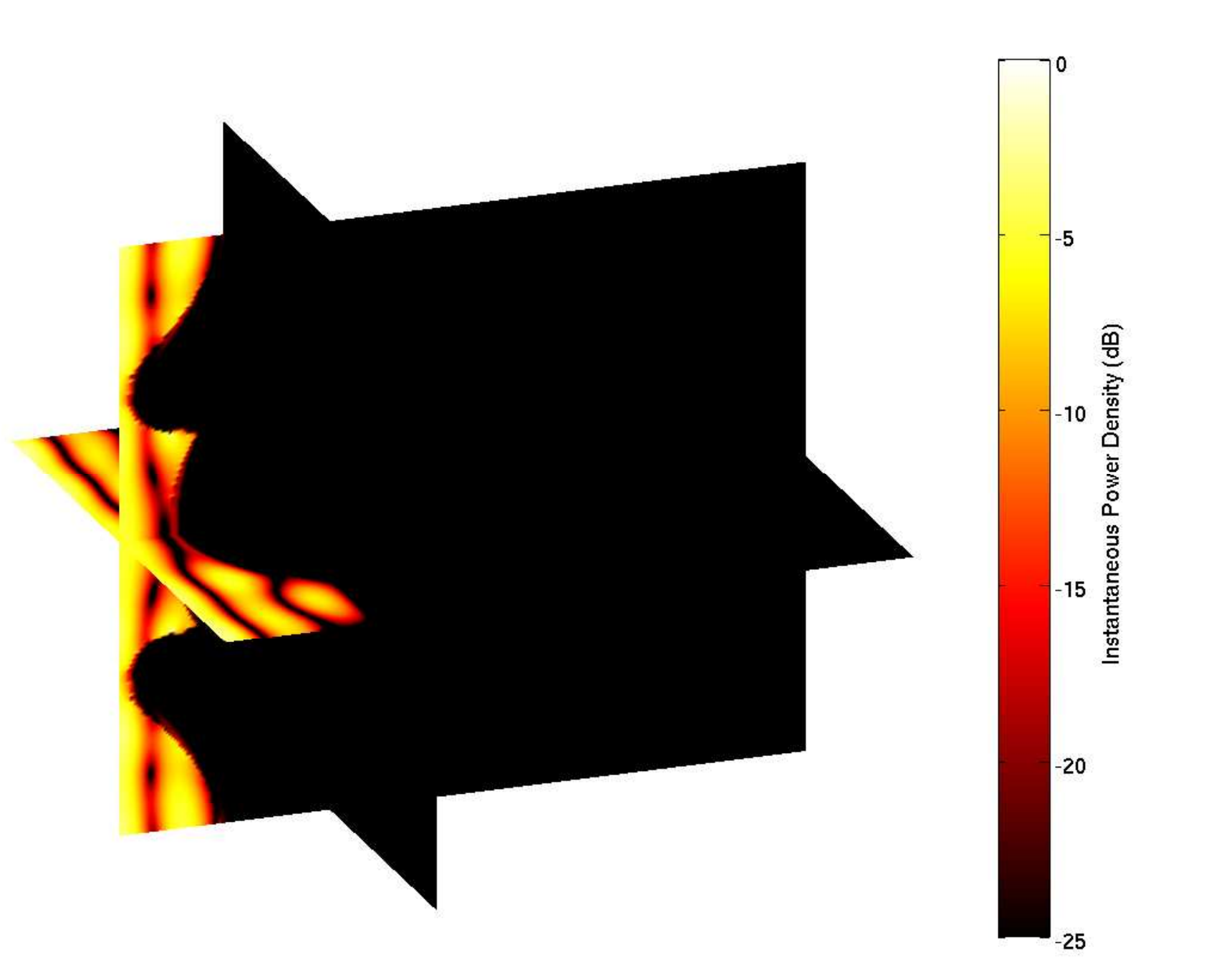}
\includegraphics[scale=0.44]{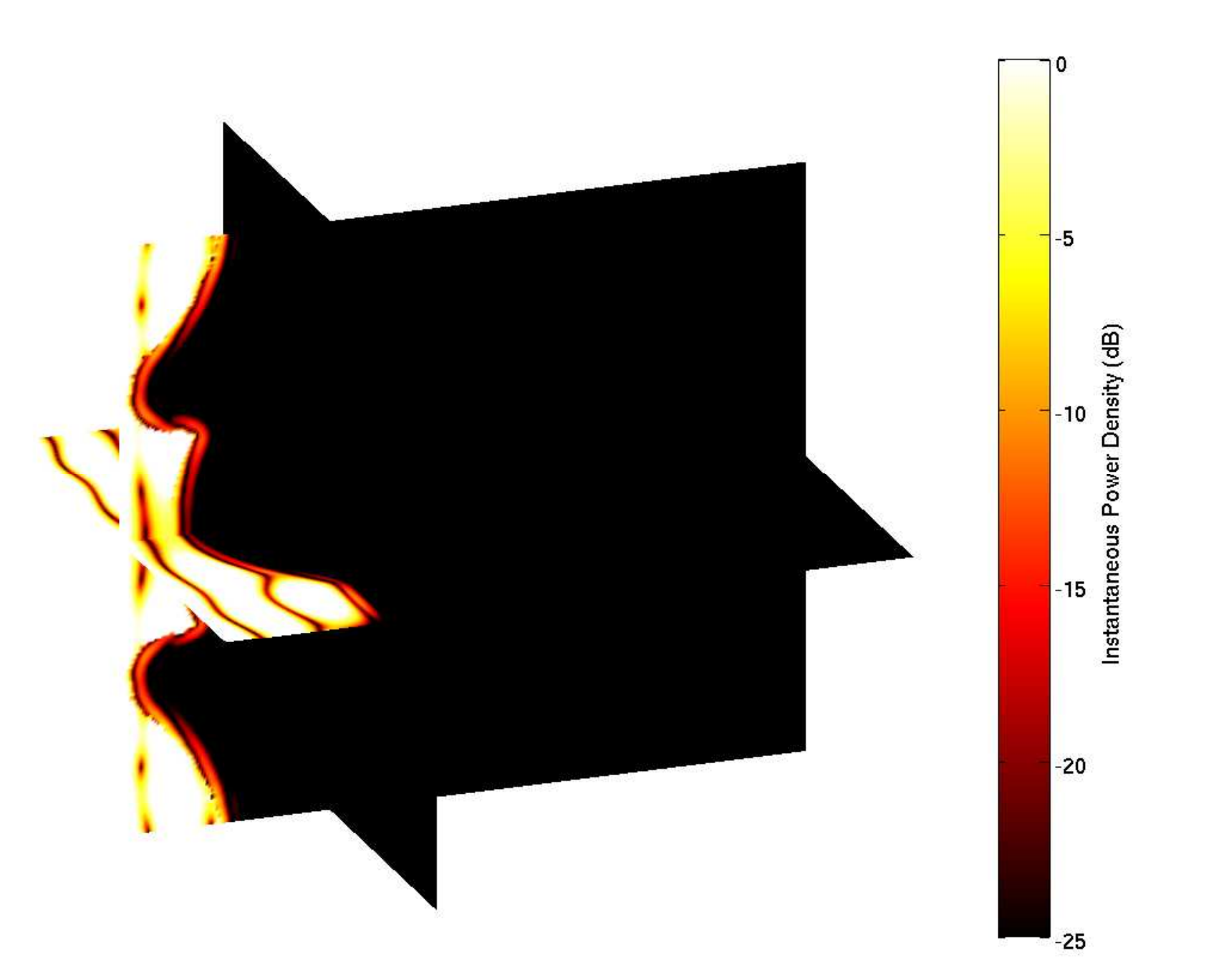}
\includegraphics[scale=0.44]{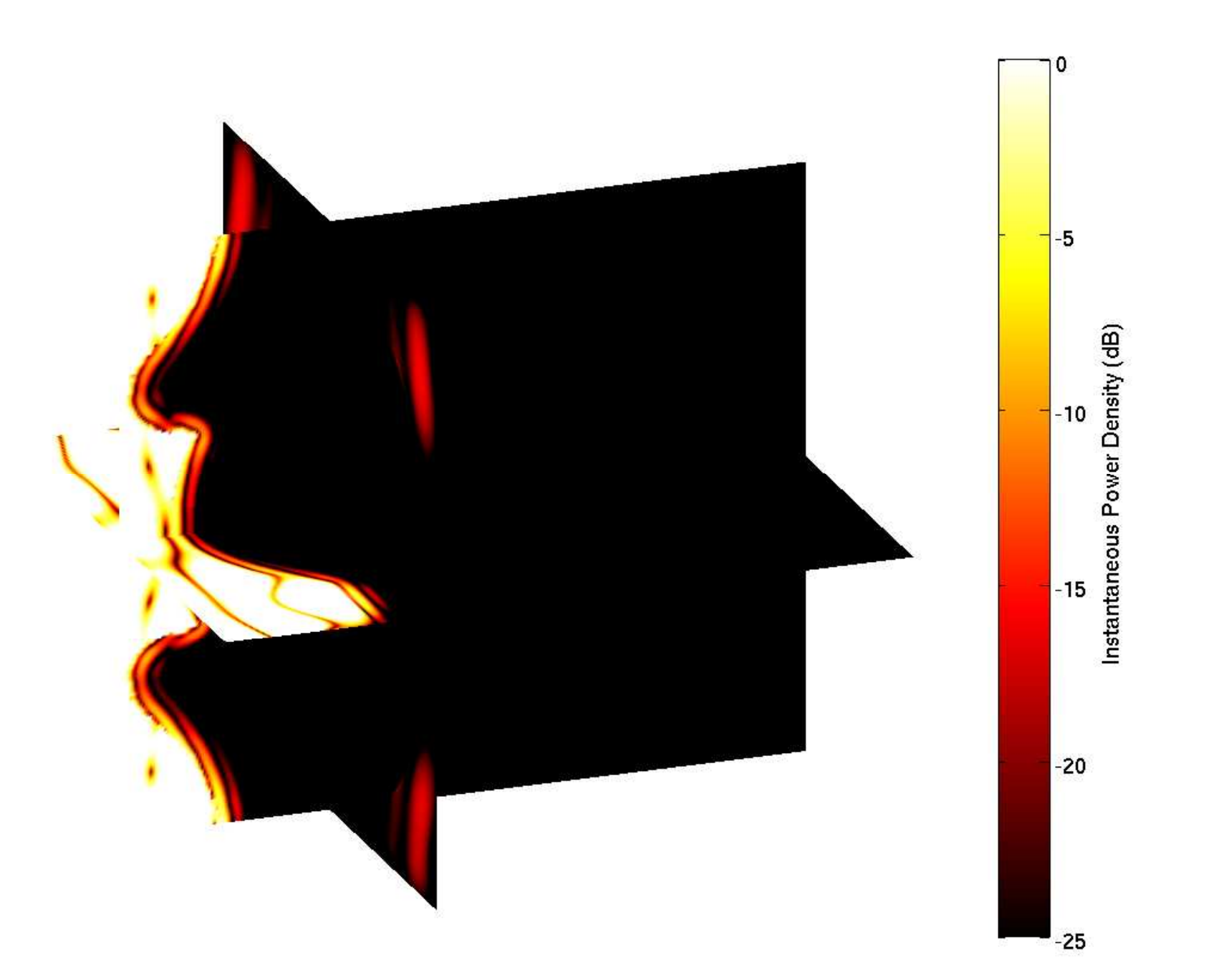}
\includegraphics[scale=0.44]{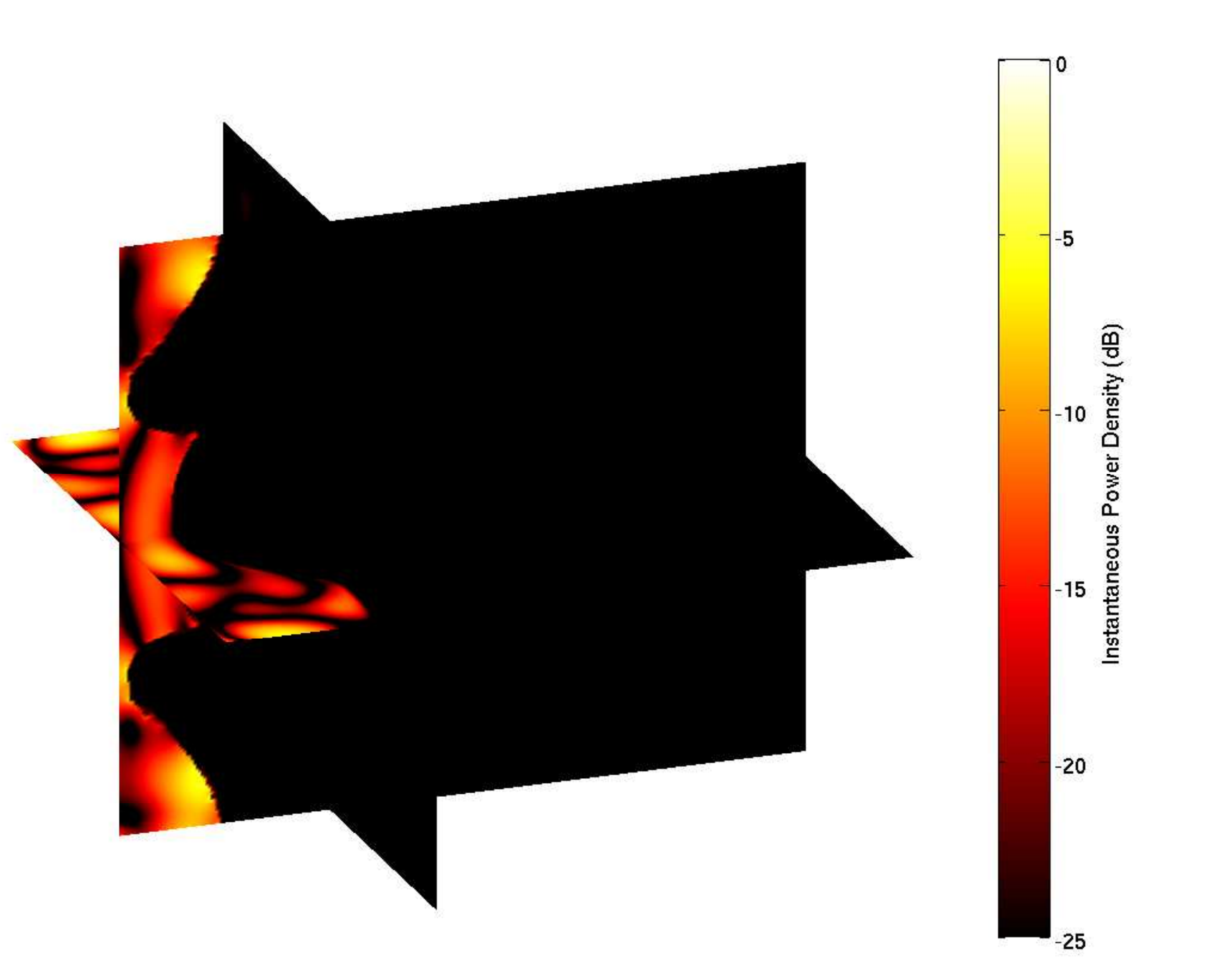}
\end{center}
\caption{Four different stages of penetration of EM pulse, in the
frequency region of $\rm 57 - 64 \; GHz$,  into the eye. The contours 
represent the instantaneous power density in decibels. Brighter regions correspond to 
higher power density.} 
\label{snaps3}
\end{figure}

A qualitative understanding of the penetration of an EM pulse into the
eye can already be achieved from the animated sequences of the pulse propagation. 
For the case of the propagation of the UWB pulses used in this paper, the 
complete animation can be accessed on-line (Simicevic 2007). Only selected  
snapshots of the intensity of the pulse penetrating the eye, shown in 
Figures \ref{snaps}, \ref{snaps2}, and \ref{snaps3}, can be presented in this paper.

To understand the dosimetric implication from the exposure of a tissue to an electromagnetic
pulse, or to model the interaction mechanism, a quantitative approach is needed.
It mostly relates to the extent of conversion of EM pulse energy into mechanical or 
thermal energy inside the tissue. A detailed study is presented in the next section.

\section{Computation of Energy Dissipation for an Electromagnetic Pulse}

If an electromagnetic pulse propagates through biological material, the 
energy will be dissipated inside the material. The conversion of electromagnetic 
energy into mechanical or thermal energy inside a volume $V$ is computed 
using equation (Jackson 1999)

\begin{equation}
P= \int_{V} \vec J \cdot \vec E  \;dV.
  \label{power1}
\end{equation}
$P$ is deposited energy per unit time, and $\vec J$ and $\vec E$ are,
respectively, the current density and electric field inside the material.  In the case of 
dispersive media, like the eye tissue, using  Equation \ref{power1} is not trivial.
First, the equation assumes a continuous distribution of charges and currents,
which is not the case for the EM pulse propagation through tissue. Second,  
the current $\vec J$ consists of the conduction or ionic currents and of the displacement currents, 
which have to be correctly resolved (Jackson 1999).

To properly calculate the amount of the absorbed energy inside the eye tissue, one first 
eliminates the current $\vec J$
in Equation \ref{power1} using the Ampere-Maxwell law (Jackson 1999). The deposited energy 
per unit time is then expressed as a function of the EM fields inside the material

\begin{equation}
P= -\int_{V} \left [ \vec \nabla \cdot (\vec E \times \vec H) 
+ \vec E \cdot {\partial \vec D \over \partial  t} 
+ \vec H \cdot {\partial \vec B \over \partial  t} \right ] \; dV.
\label{power2}
\end{equation}
$\vec D$ in this formula represents the electric displacement
and  $\vec B$  the magnetic induction. Volume integration covers the volume of the exposed 
object. 

Equation \ref{power1} consists of the term 
$\vec E \times \vec H $, representing the energy flux density, and the term 
$\vec E \cdot {\partial \vec D \over \partial  t} + \vec H \cdot {\partial \vec B \over \partial  t}$,
representing the total electromagnetic energy density. 
In the  case of linear lossless non-dispersive media, the second term is interpreted as 
the difference of internal energy per unit volume with or without the presence of the EM field.
In the case of linear dispersive media with losses, the case in this paper,  such an interpretation 
is not trivial (Landau and Lifshitz 1960). In addition, the dispersion causes temporally non-local
connection between $\vec D$  and $\vec E$ 

\begin{equation}
\vec D(\vec r,t) = \varepsilon_{0} \varepsilon_{\infty}\vec E(\vec r,t)  
+ \varepsilon_{0} \int_{0}^{t} \chi(\tau) \vec E(\vec r,t-\tau)\;d\tau.
  \label{Dt}
\end{equation}
Here $\varepsilon_{0}$ is permittivity of the free space, $\varepsilon_{\infty}$ 
the permittivity at infinite frequency, and  $\chi(\tau)$  is the susceptibility 
(Fourier transform of complex relative permittivity). This makes a direct calculation of 
total internal electromagnetic energy very difficult.

\begin{figure}
\begin{center}
\includegraphics[scale=0.5]{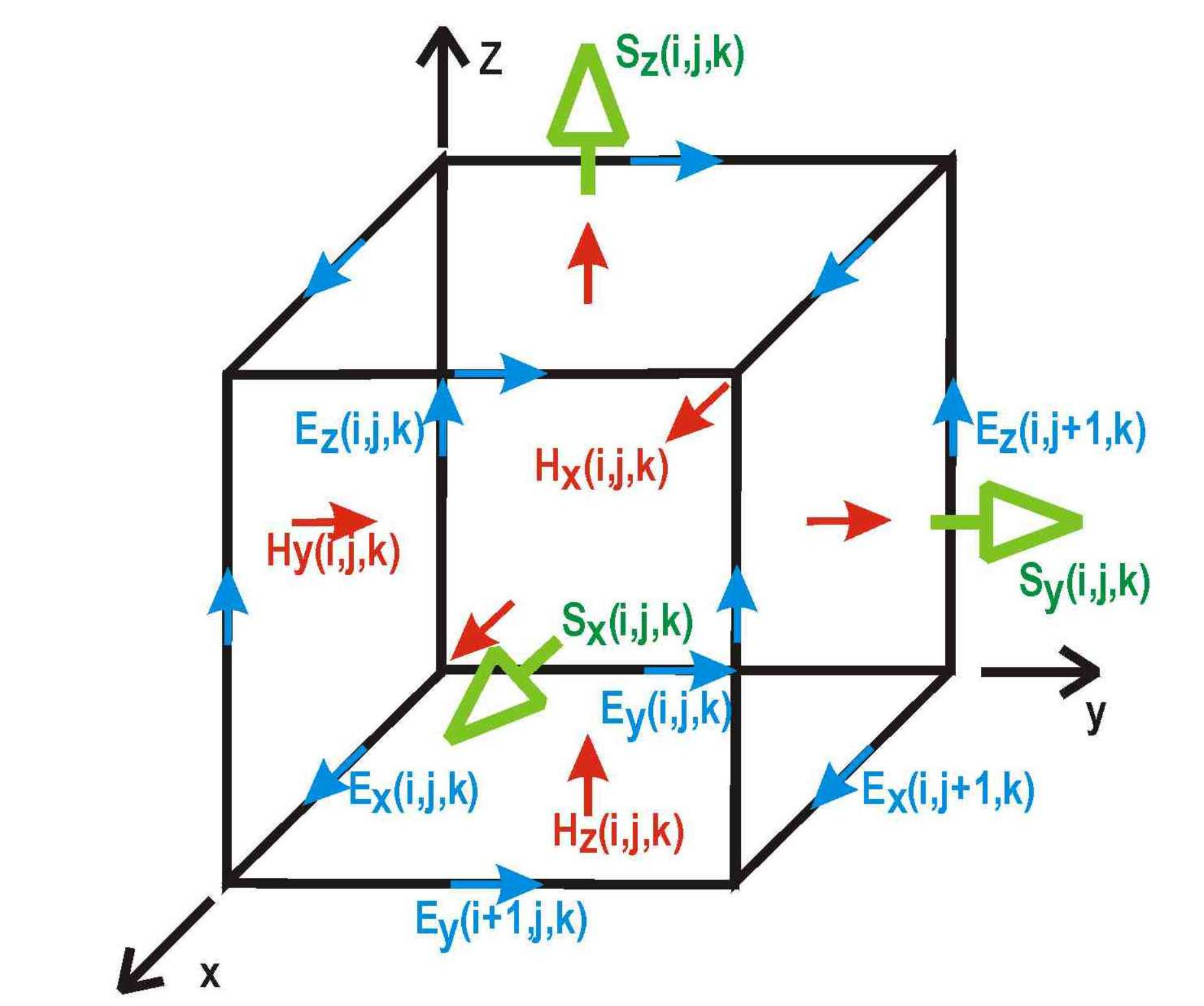}
\vspace*{- 0.5cm}
\end{center}
\caption{The placement and labeling of the EM fields and Poynting vector 
components in the Yee cell, adapted from Reference (Kunz and Luebbers 1993).}
\label{Yeecell}
\end{figure}

In this paper, the dissipation of electromagnetic energy was calculated using 
 the energy flux density. It was shown 
(Landau and Lifshitz 1960) that the formula for the energy flux density 
or Poynting vector $\vec S$

\begin{equation}
\vec S = \vec E \times \vec H ,
  \label{Poynting}
\end{equation}
is valid for variable fields even if dispersion is present. It is obvious that the difference 
of the energy flux entering a volume $V$ and the flux exiting the same volume 
represents the amount of the energy dissipated inside the volume. 
In the case of UWB radiation, this energy is obtained 
by integrating the Poynting vector over the pulse duration $T$ and the impact area $\vec A$ 
enclosing the volume

\begin{equation}
E= \int_{T}  \left [ \oint_{A} \vec S \cdot d\vec a \right ] dt.
  \label{E_Poy}
\end{equation}
Due to the absorption, this energy is ultimately converted into heat
(Landau and Lifshitz 1960). As a basic volume of interest we have chosen the Yee
cell and performed numerical integration of Equation  \ref{E_Poy} as

\begin{equation}
E= \sum_{Pulse \; duration}  \left [ \sum_{Yee \; cell \; area} \vec S \cdot \Delta \vec a \right ] \Delta t.
  \label{Poynting2}
\end{equation}
$\Delta \vec a $ is an element of the Yee cell area and $\Delta t$ is the time 
step in the FDTD computation. Due to the offset in the location of the electric and magnetic 
field components relative to each other, and relative to the position of the Poynting vector, 
the value of the Poynting vector is obtained by using the average values of the adjacent 
field components. As an example, with the
notation from Figure \ref{Yeecell}, the $S_{x}$ component of the Poynting vector $\vec S$ at a
position $(I,J,K)$ and time $t$ is calculated as
\begin{eqnarray}
S_{x}(I,J,K)  =  {{E_{y}(I,J,K)+E_{y}(I,J,K\!+\!1)} \over 2} \nonumber          \\
                        \times  {{H_{z}(I,J,K)+H_{z}(I,J,K\!+\!1)+H_{z}(I\!-\!1,J,K)+H_{z}(I\!-\!1,J,K\!+\!1)} \over 4}  \nonumber \\
                         - {{E_{z}(I,J,K)+E_{z}(I,J\!+\!1,K)} \over 2}   \nonumber \\
                        \times {{H_{y}(I,J,K)+H_{y}(I,J\!+\!1,K)+H_{y}(I\!-\!1,J,K)+H_{y}(I\!-\!1,J\!+\!1,K)} \over 4}.
  \label{Sxnum}
\end{eqnarray}
Components $S_{y}$ and $S_{z}$ are calculated in a similar way.

The numerical accuracy of Equation \ref{Poynting2} was tested for the case of an
EM pulse propagating through empty space where the deposited energy 
inside the Yee cell should be $0$.  In the worst case, the accuracy of the method, 
measured as the  deviation from $0$,  was found to be less than 0.02 \%, 
a negligible error compared, for example, 
to the accuracy of the dielectric properties of tissues. 

Numerical computation of the absorbed energy was also 
validated by comparing the results obtained using Equation \ref{Poynting2} with
those obtained by analytical solution. For a normal incidence of an EM
wave on a infinite conducting surface of a dielectric constant 
$\varepsilon = \varepsilon_{R} - i \varepsilon_{I}$, the transmitted energy flux or transmittance
can be calculated as

\begin{equation}
T= 1-{{(\sqrt\varepsilon_{R} -1)^{2} + \varepsilon_I} \over {(\sqrt\varepsilon_{R} +1)^{2} + \varepsilon_I} }
  \label{transmittance}
\end{equation}
The dielectric constant  for a muscle at a frequency of $\rm 7 \; GHz$, 
the mid frequency of  $\rm 3.1 - 10.6 \; GHz$ range pulse, is $\varepsilon = 46.9 - i 16.6$. 
For this dielectric constant, Equation \ref{transmittance} gives for the transmittance a value of $35 \%$. 
Equation \ref{Poynting2} gives for the transmittance of a $\rm 3.1 - 10.6 \; GHz$ range UWB pulse 
$33 \%$.  This is as good as expected agreement taking into consideration that the 
absorbed energy is computed only to the depth of  $\rm 20 \; mm$  and that the transmittance 
of the UWB pulse was compared to transmittance of a plane wave having the pulse's mid frequency.

\section{Dissipation of Electromagnetic Pulse Energy Inside the Eye}

\begin{figure}
\begin{center}
\includegraphics[scale=0.5]{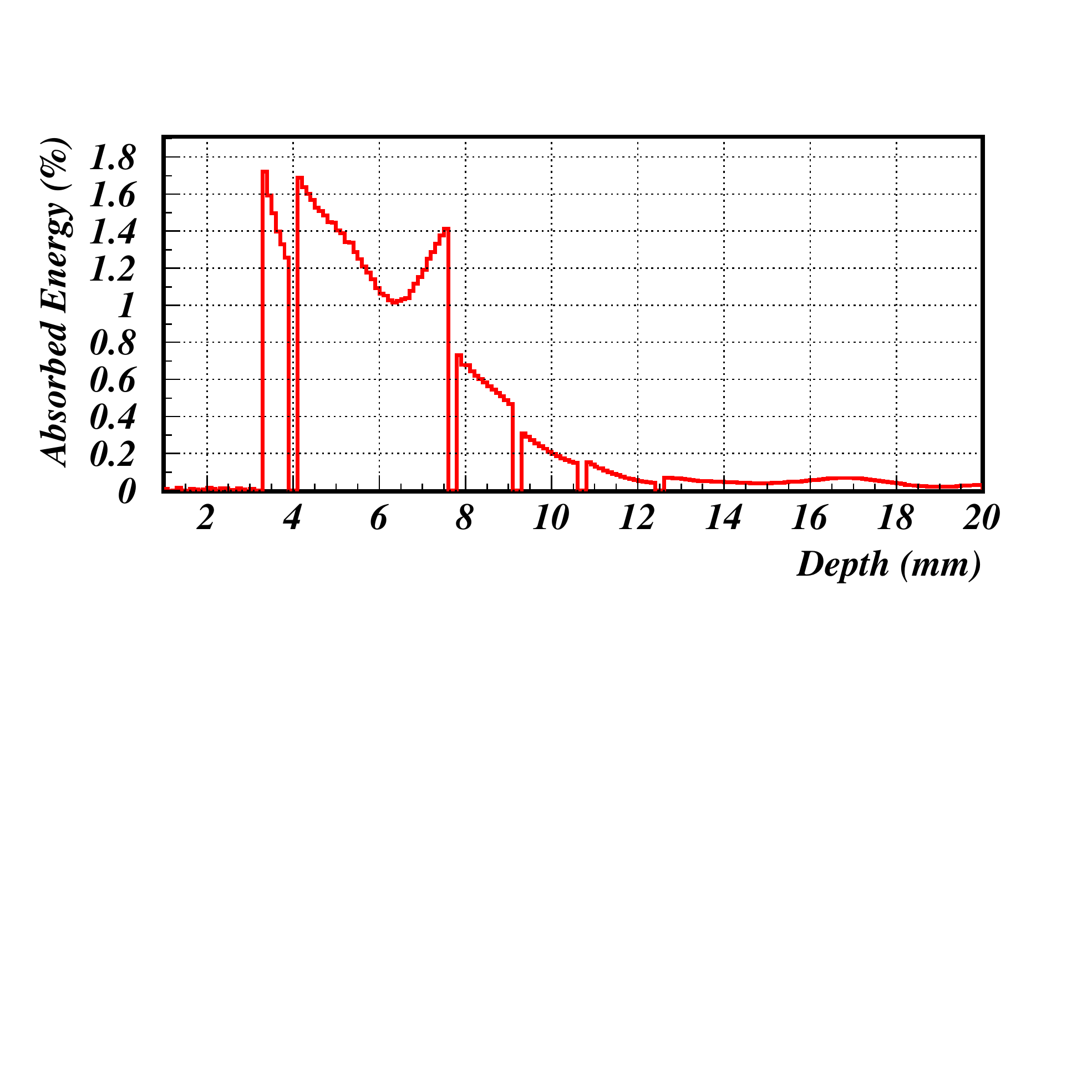}
\end{center}
\begin{center}
\vspace*{- 5.5cm}
\includegraphics[scale=0.5]{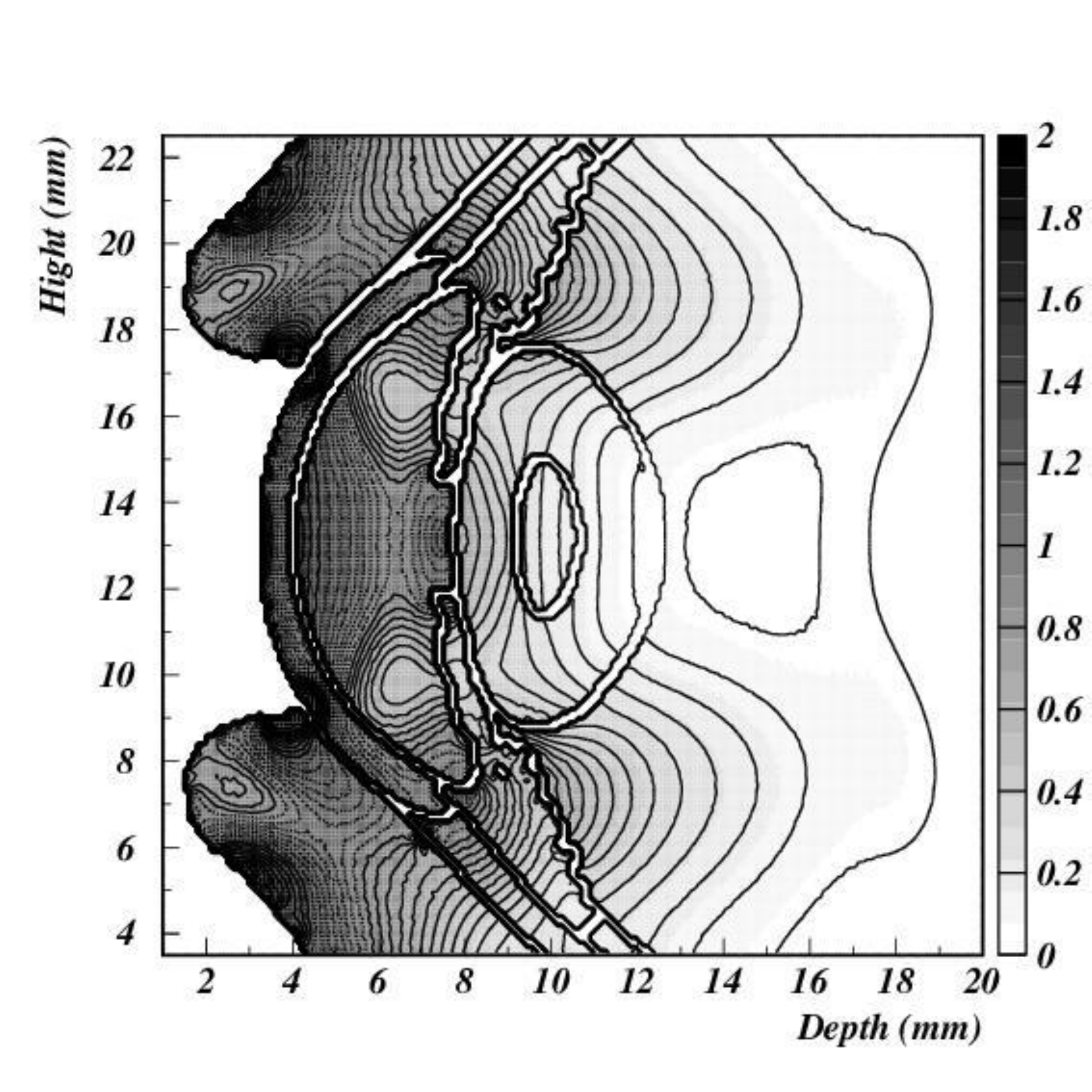}
\vspace*{- 1.cm}
\end{center}
\caption{Contour plot of the density of the energy absorption across the vertical cross section of the
eye in percentage of the total available energy for the $\rm 3.1 - 10.6 \; GHz$ range UWB pulse. 
Distribution of the absorbed energy along the mid-line is shown in the top figure. From left to right 
separate regions corespond to cornea, anterior chamber, lens cortex, lens nucleus, lens cortex (again),
and vitreous humor.}
\label{en_absorb_3_10}
\end{figure}

\begin{figure}
\begin{center}
\includegraphics[scale=0.35]{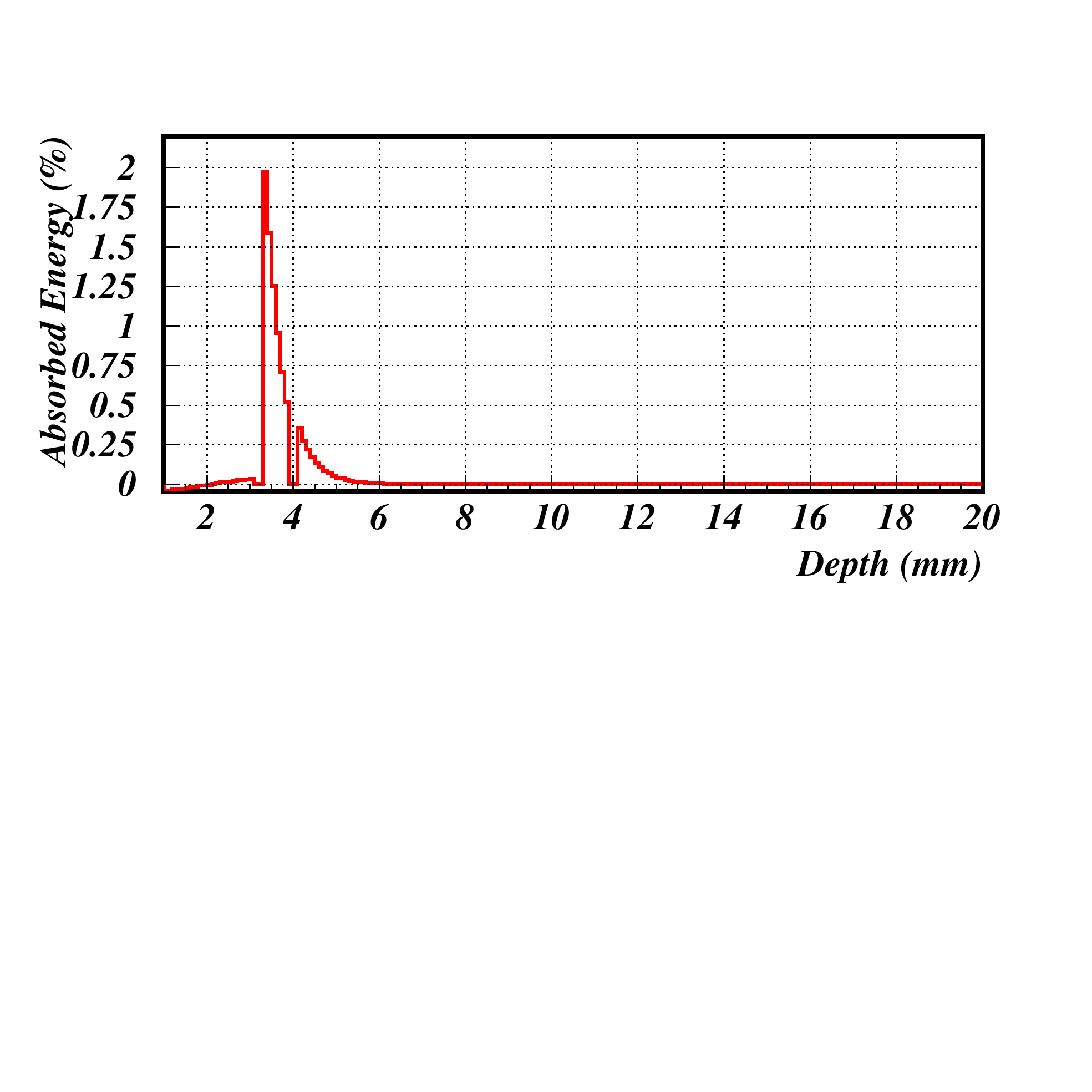}
\includegraphics[scale=0.35]{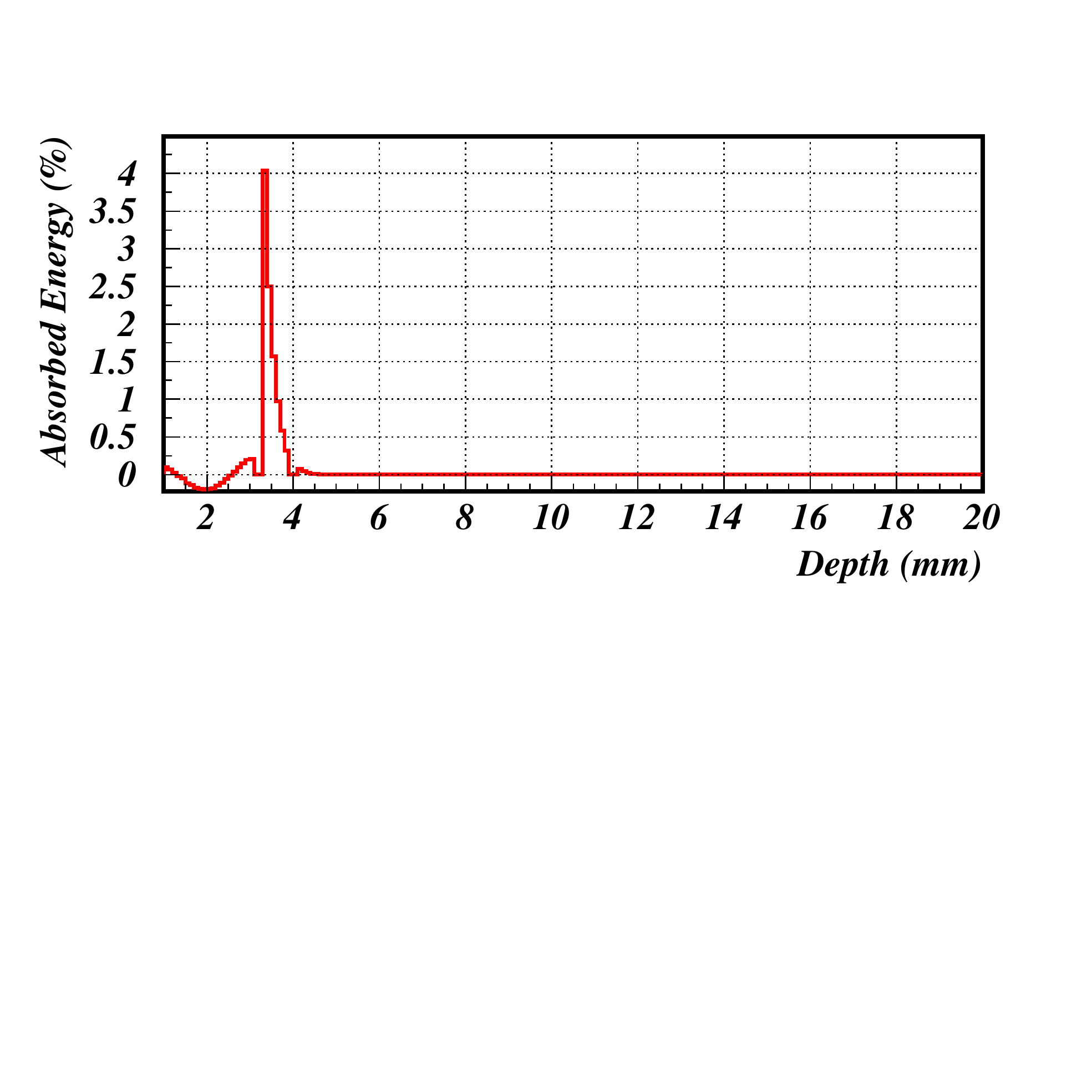}
\end{center}
\vspace*{- 4.cm}
\begin{center}
\includegraphics[scale=0.35]{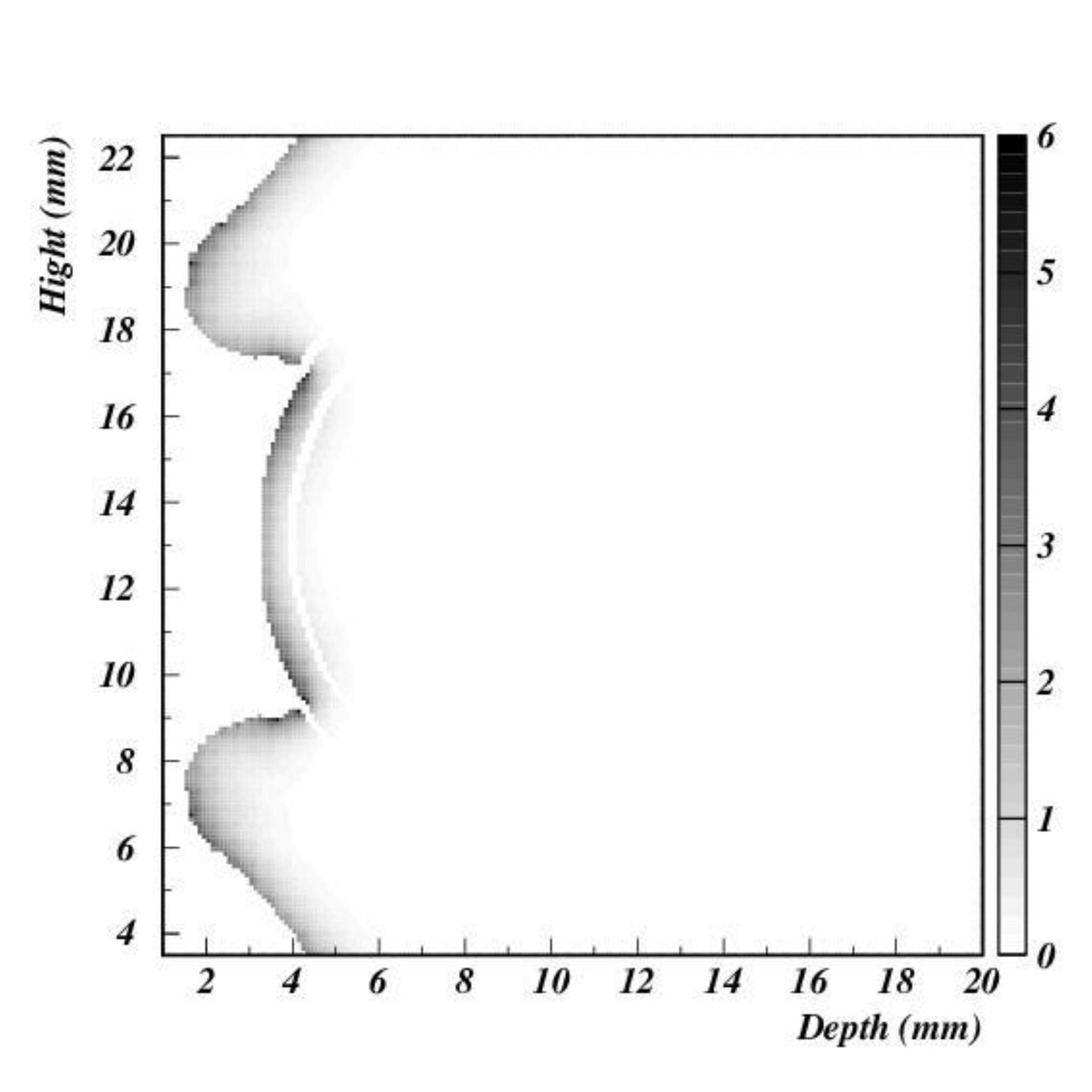}
\includegraphics[scale=0.35]{figure10c.pdf}
\end{center}
\caption{Contour plots of the density of the energy absorption across the vertical cross section of the
eye in percentage of the total available energy for the $\rm 22 - 29 \; GHz$ (left) 
and $\rm 57 - 64 \; GHz$ (right) frequency range UWB pulses. 
Distributions of the absorbed energy along the mid-line are shown on the top. From left to right 
separate regions corespond to cornea and anterior chamber.}
\label{en_absorb_22_64}
\end{figure}

The penetration of the EM pulses into a human eye is shown in  Figures \ref{en_absorb_3_10} 
and \ref{en_absorb_22_64}. As expected, the penetration depth decreases as the pulse's 
frequency spectrum increases. In the case of the pulse in the lowest frequency range,
as shown in Figure \ref{en_absorb_3_10}, 
significant energy penetrates  about $\rm 10 \; mm$ into the eye where it is, as a result
of the complex eye structures,  relatively nonhomogeneously absorbed.  
Surprisingly significant contribution to energy absorption modulation 
comes from, in computation, typically neglected structures as, for example, the iris. Also, 
unexpectedly, most of the energy in the cornea is absorbed next to  the eye lids. 
The $\rm 22 - 29 \; GHz$ range UWB pulse significantly penetrates into the eye only
$\rm 2 - 3 \; mm$ (Figure \ref{en_absorb_22_64}). Most of the energy is absorbed by the 
cornea with the energy absorption hot spots again immediately above or below the eye lids. 
The  $\rm 57 - 64 \; GHz$ range UWB pulse is almost entirely and uniformly absorbed by the cornea. 
While the energy absorption data exist for the entire eye volume, only selected data can be 
presented in this paper.
 
Most of the experimental work on the effects of eye exposure to electromagnetic radiation 
in the frequency range covered in this paper was performed using continuous waves. We will
compare the results of the UWB pulse propagation with the results of the CW, restricting ourselves to the 
frequencies above $\rm 10 \; GHz$. Above $\rm 10 \; GHz$ experimental results 
exist for the exposure to $\rm 35 \; GHz$ waves (Rosenthal {\it et al} 1977, Chalfin {\it et al} 2002) and
$\rm 60 \; GHz$ waves  (Kues {\it et al} 1999, Kojima {\it et al} 2005). A short but detailed summary of the  
experiments, including the experimental apparatus, organic material, cellular environment, 
and possible electromagnetic interaction mechanisms can be found in Miller {\it et al} (2002).

\begin{figure}
\begin{center}
\includegraphics[scale=0.5]{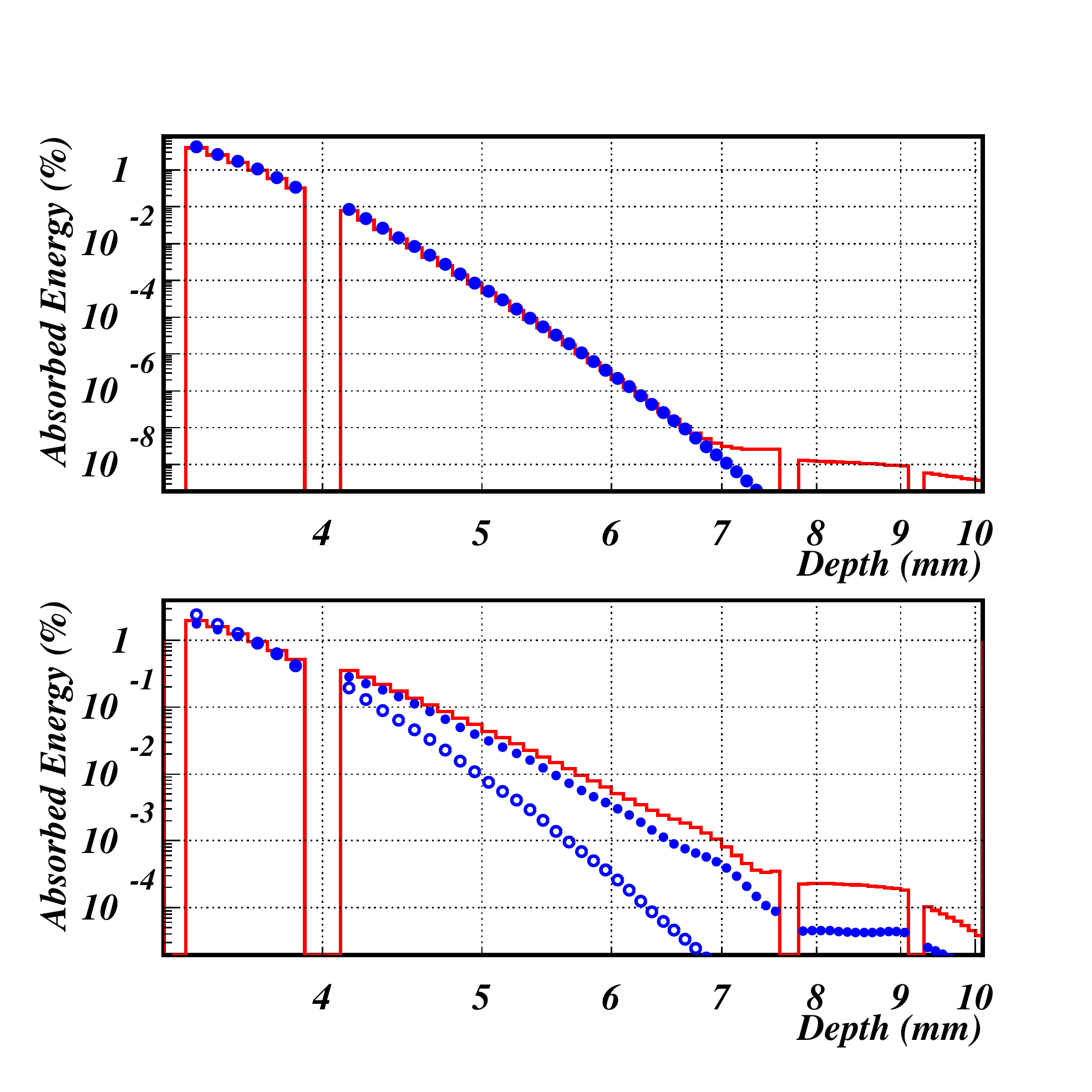}
\caption{\label{en_compar} Top: comparison of the absorbed energy after exposure
to $\rm 57 - 64 \; GHz$ frequency range UWB pulse (red line) and after exposure
to $\rm 60 \; GHz$ CW (blue circles). Bottom: comparison of the absorbed energy after exposure
to $\rm 22 -  29 \; GHz$ frequency range UWB pulse (red line) and after exposure
to $\rm 22.5 \; GHz$ (filled blue circles) and $\rm 35 \; GHz$ CW (hollow blue circles). 
From left to right 
separate regions corespond to cornea, anterior chamber, lens cortex, and lens nucleus.} 
\end{center}
\end{figure}

Searching for any abnormal effects resulting from the eye exposure to UWB radiation,  
the computed distribution of the absorbed energy in the case of UWB exposure was 
compared to the energy distribution in the case of CW.  
As shown in Figure \ref{en_compar}, no difference in the amount 
of absorbed energy was found when  the $\rm 57 - 64 \; GHz$ frequency range UWB pulse 
was compared to $\rm 60 \; GHz$ CW, or when the $\rm 22 - 29 \; GHz$  frequency range UWB pulse 
was compared to $\rm 22.5 \; GHz$ CW, both corresponding to mid-frequency of the pulses spectra. 
 When $3\rm 5 \; GHz$  CW was compared to a $\rm 22 - 29 \; GHz$  frequency range pulse
(bottom of  Figure \ref{en_compar}), as expected, the 
pulse penetrated deeper into the eye due to its lower frequency spectrum. 
No extraordinary effects outside the consequences from the Fourier decomposition of the UWB pulse 
were observed. Two factors played a role in the energy 
dissipation of the UWB electromagnetic radiation: the frequency spectrum of the pulse and the 
dielectric properties of the exposed biological material at those frequencies.

The consequences of this study is summarized in the Table \ref{tab2}. As was shown,  
the energy in the material exposed to  UWB electromagnetic radiation was absorbed 
in the same way as in the material exposed to a CW in the coresponding pulse frequency spectrum. 
The absorbed energy in the tissues was, in both cases, and in the same way, proportional to the 
total available EM energy. 

The total available energy in the case of CW can be easily calculated from the Poynting vector
\begin{equation}
S = {c\varepsilon_{0} \over 2}  E_{0}^{2},
  \label{CW_poynting}
\end{equation}
where $c$ is the speed of light, $ \varepsilon_{0}$ the vacuum permittivity, and $E_{0}$ the CW
electric field amplitude. The energy of the UWB pulse can be easily obtained by integration 
\begin{equation}
S = c\varepsilon_{0}  \int_{T} E(t)^{2} dt
  \label{CW_poynting2}
\end{equation}
over the pulse time duration $T$. $E(t)$ is the electric field at the time $t$. It is easy to 
see that, for the same field amplitude, the energy carried by the short EM pulses
depends on the pulse repetition rate and it is always 
much smaller than the energy carried by the CW. The results are summarized in Table \ref{tab2}. 
Assuming the same exposure time, for the pulse repetition rate of 1MHz, the energy absorbed 
by the exposed material will be $\sim$ 0.01 \% of the corresponding CW exposure, for the 
maximum repetition rate of ~1 GHz, the energy absorbed will be $\sim$ 10 \% of the 
coresponding CW energy.  Any possible absorbed-dose-related health effects as a results 
of the exposure to UWB will be 
therefore reduced by one to many orders of magnitude compared to the health effects 
caused by the exposure to CW. To quote from Miller {\it et al} (2002):
`` We want to close this section with a more general observation on the effects of UWB pulses.
When we first started this work, there was speculation that even one UWB pulse might have 
serious effects on biological organisms. So far, we have exposed animals to over a quarter of
a billion UWB pulses, and we have never seen any acute effects."

\begin{table}
\begin{center}
\begin{tabular}{ccccccc}
\multicolumn{1}{c}{Pulse Frequency Range (GHz)} {\vline}&
\multicolumn{1}{c}{RepRate: 1 MHz} {\vline}&
\multicolumn{1}{c}{RepRate: 1 GHz} {\vline}\\
\hline\hline
  $3.1 - 10.6 $  & 0.011 \% & 11\%  \\
 $22 - 29 $  & 0.009 \% & 9  \% \\
 $57 - 64 $  & 0.009 \% & 9  \% \\
\end{tabular}
\end{center}
\caption{ Energy carried by the UWB pulse as a percentage of the energy carried by the
CW of the same field amplitude, and in the same time interval. In one case $10^{6}$ and in the other
$10^{9}$ pulses were generated in one second.}
\label{tab2}
\end{table}

\section{Conclusion}

In this paper, we have performed full three dimensional FDTD calculations 
of the penetration of UWB electromagnetic pulses, authorized 
by the FCC for communications, radar imaging, and  vehicle radars,
into a human eye. Calculations included as  detailed geometrical description of the eye as 
necessary and as accurate description 
of the physical properties of the eye tissue as allowed by existing data. 
The spatial resolution of $ \rm 0.1 \; mm$ side length of the Yee cell allowed 
reliable calculation of up to $\rm \sim 90 \; GHz$ in frequency range in the dielectric.
 
To minimize the computation time, the electromagnetic interaction with dielectric 
material was modeled using the Piecewise-Linear Recursive Convolution 
method (PLRC) for Debye media. Dielectric 
properties of the eye tissues in the frequency range $\rm \leq 100 \; GHz$ were formulated  
in terms of the Debye parametrization with the same accuracy as 
the accepted Cole-Cole parametrization.

The energy absorbed after the UWB  exposure was evaluated over the entire eye volume
and compared to the energy absorbed after the exposure to continuous 
electromagnetic waves.  The energy in the material exposed to  UWB electromagnetic 
radiation was absorbed in the same way as in the material exposed to a 
CW in the coresponding pulse frequency spectrum. We have found that,
assuming the same field amplitude and the same exposure time,  
any possible dose dependent health effects as a results of the exposure to UWB 
radiation will be reduced by one to many orders of magnitude compared to the 
health effects caused by the exposure to CW. This finding is in agreement
with the experiments carried out at the  US Air Force Research Laboratory at Brooks
 (Miller {\it et al} 2002). 

We can conclude that, based on the research described in this paper, any future applications 
of UWB pulses, for the purposes and in the frequency range approved by the FCC, pose less health risk
compared to those applications being carried by the continuous electromagnetic radiation.

\section*{Acknowledgments}

I would like to thank Jadranka Simicevic, Raymond L Sterling and Steven P Wells 
for helpful suggestions.
 
\section*{References}

\begin{harvard}

\item[] Barnes F S  and Greenebaum B eds. 2006 {\it Handbook of Biological Effects of
Electromagnetic Fields - third edition} Boca Raton: CRC Press LLC.
\item[] Chalfin S, D’Andrea J A, Comeau P D, Belt M E and Hatcher D J 2002  
{\it Health Physics} {\bf 83} 83
\item[] Charles M W and Brown N 1975 {\it Phys. Med. Biol.}  {\bf 20} 202
\item[] de AlmeidaI M S and Carvalho L A 2007 {\it Braz. J. Phys.}  {\bf 37} 378
\item[] Federal Communications Commission 2002, News Release NRET0203, 
http://www.fcc.gov
\item[] Gabriel C 1996 {\it Preprint} AL/OE-TR-1996-0037,
Armstrong Laboratory Brooks AFB
\item[] Gabriel C, Gabriel S, and Corthout E 1996 {\it Phys. Med. Biol.}  {\bf 41}  2231
\item[] Gabriel S, Lau R W  and Gabriel C 1996 {\it Phys. Med. Biol.}  {\bf 41}  2251
\item[] Gustavsen B and Semlyen A 1999 {\it IEEE Trans. Power Delivery} {\bf 14} 1052
\item[] Hu Q, Viswanadham S, Joshi R P, Schoenbach K H, Beebe S J 
and Blackmore P F 2005 {\it  Phys. Rev. E} {\bf 71} 031914
\item[] ICNIRP 1998 {\it Health Physics} {\bf 74} 494
\item[] IEEE Standards Coordinating Committee 28 on Non-Ionizing Radiation Hazards: 
Standard for safety levels with respect to human exposure to radio frequency electromagnetic fields, 
3 kHz to 300 GHz (ANSI/IEEE C95.1-1991), The Institute of Electrical and Electronics Engineers, 
New York; 1992.
\item[] Jackson J D 1999 {\it Classical Electrodynamics} New York:John Willey \& Sons Inc.
\item[] Ji Z, Hagness S C, Booske J H, Mathur S, and Meltz M 2006 
{\it IEEE Transactions on Biomedical Engineering} {\bf 53} 780
\item[] Kojima M, Yamashiro Y, Hanazawa M, Sasaki H, Wake K, Watanabe S, Taki M, 
Kamimura Y and Sasaki K 2005 {\it Proceedings of the XXVIIIth URSI General 
Assembly in New Delhi}  K02.9(470) 
\item[] Kues H, D’Anna S A, Osiander R, Green W R and  Monahan J C 1999 
{\it Bioelectromagnetics}  {\bf 20} 463 
\item[] Kunz K and Luebbers R 1993 {\it  The Finite Difference Time Domain Method for
Electromagnetics} Boca Raton: CRC Press LLC.
\item[] Landau L D and Lifshitz 1960 {\it Electrodynamics of Continuous Media} Reading:
Addison-Wesley Inc.
\item[] Lotmar W 1971 {\it J. Opt. Soc. Am. }  {\bf 61} 1522
\item[] Luebbers R J, Hunsberger F, Kunz K S, Standler R B, and Schneider M 1990
 {\it IEEE Trans. Electromagn. Compat.} {\bf 32} 222
\item[] Luebbers R J, Hunsberger F, and Kunz K S 1991
{\it IEEE Trans. Antennas Propagat.} {\bf 39} 29
\item[] Luebbers R J and Hunsberger F 1992 {\it IEEE Trans. Antennas Propagat.} 
{\bf 40} 1297
\item[] Lugo G R N 2006  {\it Modelos dispersivos para analisis de la propagacion 
de campos electromagneticos en tejidos biologicos} Master thesis, Universidad Autonoma de 
Nuova Leon, Mexico
\item[] Miller R L, Murphy M R and Merritt J H 2002 
{\it Proceding of the 2nd International Workshop on Biological Effects of EMFs} 
Rhodes, Greece; 2002:468-477.
\item[] Norrby S, Piers P, Campbell C and van der Mooren M 2007 {\it Applied Optics} 
accepted for publication
\item[] Polk C and Postow E eds. 1995 {\it Handbook of Biological Effects of
Electromagnetic Fields - second edition} Boca Raton: CRC Press LLC.
\item[] Rosenthal S W, Birenbaum L, Kaplan I T, Metlay W, Snyder W Z and Zaret M M 1977
{\it HEW Publication (FDA)} {\bf 77-8010} 110
\item[] Sadiku M N O 1992 {\it  Numerical Techniques in Electromagnetics}
Boca Raton: CRC Press LLC.
\item[] Schoenbach K H, Joshi R P, Kolb J F, Chen N, Stacey M, Blackmore P F,
Buescher E S, and Beebe S J 2004 {\it Proceedings of the IEEE} {\bf 92} 1122
\item[] Siedlecki D, Kasprzak H and Pierscionek B K 2004 {\it Optics Letters}  {\bf 29} 1197
\item[] Simicevic N and Haynie D T 2005 {\it Phys. Med. Biol.}  {\bf 50} 347
\item[] Simicevic N 2005 {\it Phys. Med. Biol.}  {\bf 50} 5041
\item[] Simicevic N 2007 {\it Health Physics. The Radiation Safety Journal} {\bf 92} 574 
\item[] Simicevic N 2007 {\it http://caps.phys.latech.edu/$\sim$neven/pmb/}
\item[] Sullivan, D M 2000 {\it Electromagnetic Simulation Using the FDTD Method }
New York: Institute of Electrical and Electronics Engineers.
\item[] Taylor J D ed. 1995 {\it Introduction to Ultra-Wideband Radar Systems}
Boca Raton: CRC Press LLC.
\item[] Taflove A and Hagness S C 2000  {\it  Computational Electrodynamics:
The  Finite-Difference  Time-Domain  Method,  2nd  ed.  }  Norwood:  Artech  House.
\item[] Yee K S 1966 {\it IEEE Trans. Antennas Propagat.} {\bf AP-14} 302
\item[] Zastrow E, Davis S K and Hagness S C 2007 
{\it  Microwave and Optical Technology Letters} {\bf 49}  221

\end{harvard}

\end{document}